\documentclass[twocolumn]{aastex701}
\shorttitle{PANGU I: Fixed-Spacetime GRMHD}
\shortauthors{Li et al.}

\usepackage[T1]{fontenc}
\usepackage{amsmath,amssymb,bm}
\usepackage{booktabs}
\usepackage{multirow}
\usepackage{graphicx}
\usepackage{hyperref}
\usepackage{placeins}
\usepackage{xcolor}
\usepackage{tcolorbox}
\usepackage{etoolbox}

\hypersetup{colorlinks=true,citecolor=blue!55!black,linkcolor=blue!55!black,urlcolor=blue!55!black}
\newcommand{\dd}{\mathrm{d}}

\newcommand{\pangu}{\texttt{PANGU}}
\newcommand{\rg}{r_{\rm g}}
\newcommand{\tg}{t_{\rm g}}
\newtcbox{\figframe}[1][]{on line,colback=white,colframe=black!55,boxrule=1.05pt,arc=2mm,boxsep=0pt,left=1.1mm,right=1.1mm, top=1.1mm,bottom=1.1mm,#1}

\makeatletter
\newcommand{\settablecaptype}{\def\@captype{table}}
\makeatother
\begin{document}

\title{\texttt{PANGU} I: A \texttt{Parthenon}-Based Framework for Fixed-Spacetime GRMHD}

\correspondingauthor{Minyong Guo}
\email{minyongguo@bnu.edu.cn}

\correspondingauthor{Bin Chen}
\email{chenbin1@nbu.edu.cn}

\author[0009-0007-4339-0570]{Yuehang Li}
\email{liyuehang26@stu.pku.edu.cn}
\affiliation{School of Physics and Astronomy, Beijing Normal University, Beijing 100875, People's Republic of China}
\affiliation{School of Physics, Peking University, No.5 Yiheyuan Rd, Beijing 100871, People's Republic of China}

\author[0009-0001-0796-1547]{Fan Zhou}
\email{202631101012@mail.bnu.edu.cn}
\affiliation{School of Physics and Astronomy, Beijing Normal University, Beijing 100875, People's Republic of China}

\author[0009-0007-2703-6221]{Mingyang Luan}
\email{202521101021@mail.bnu.edu.cn}
\affiliation{School of Physics and Astronomy, Beijing Normal University, Beijing 100875, People's Republic of China}

\author[0000-0001-5577-575X]{Minyong Guo}
\email{minyongguo@bnu.edu.cn}
\affiliation{School of Physics and Astronomy, Beijing Normal University, Beijing 100875, People's Republic of China}
\affiliation{Key Laboratory of Multiscale Spin Physics, Ministry of Education, Beijing, 100875, People's Republic of China}

\author[0000-0003-4509-9705]{Bin Chen}
\email{chenbin1@nbu.edu.cn}
\affiliation{School of Physics, Peking University, No.5 Yiheyuan Rd, Beijing 100871, People's Republic of China}
\affiliation{Institute of Fundamental Physics and Quantum Technology, \& School of Physical Science and Technology, Ningbo University, Ningbo, Zhejiang 315211, People's Republic of China}
\affiliation{Zhejiang Key Laboratory of Extreme Universe, \& BINGO Center, Ningbo University, Ningbo, Zhejiang 315211, \\People's Republic of China}

\begin{abstract}
Existing general-relativistic magnetohydrodynamics (GRMHD) codes offer complementary capabilities, but these capabilities remain distributed across different frameworks.
To unify them, we introduce \pangu{}---\textbf{P}arthenon-Based \textbf{A}strophysics for \textbf{N}umerical Relativity and \textbf{G}RMHD in a \textbf{U}nified Framework---a modular computational framework that integrates a shared solver across physical regimes, flexible spacetime geometry, and extensible computational infrastructure within a single architecture. In the first paper of this series, we present and validate the fixed-spacetime GRMHD component of \pangu{}, in which the spacetime geometry is prescribed and is not dynamically evolved in response to the matter stress-energy tensor.
The shared solver combines finite-volume Godunov methods with constrained transport on uniform and block-structured, statically refined meshes, and covers hydrodynamics and magnetohydrodynamics in Newtonian, special-relativistic, and general-relativistic regimes.
\pangu{} provides a flexible geometry interface supporting coordinate selection and either on-the-fly or precomputed evaluation of spacetimes, demonstrated here in Kerr spacetime with Cartesian Kerr--Schild and modified Kerr--Schild coordinates; additional components can be incorporated without restructuring the core solver.
Verification covers all three regimes on uniform and statically refined meshes: smooth Newtonian and special-relativistic problems approach second order against their analytic solutions, with deliberately matched double-precision trajectories agreeing with \texttt{AthenaK} to roundoff; in fixed-spacetime GRMHD, the stationary analytic solutions are preserved to second order, and the turbulent accretion reproduces the same flow and flux scales as \texttt{BHAC}.
Together, these results establish the fixed-spacetime matter-sector foundation for the broader \pangu{} framework.
Numerical-relativity and coupled dynamical-spacetime GRMHD capabilities will be presented in subsequent work.
\end{abstract}


\section{Introduction}
\label{sec:introduction}

General-relativistic magnetohydrodynamics (GRMHD) provides the standard framework for modeling magnetized relativistic flows around compact objects, including black-hole accretion disks, relativistic jets, and compact-object mergers \citep{Gammie2003HARM,Porth2019CodeComparison}.
Numerical solutions of the GRMHD equations must simultaneously capture shocks and relativistic wave propagation, maintain the solenoidal magnetic-field constraint, recover physical primitive states in highly magnetized regions, and consistently couple the fluid evolution to curved spacetime geometry.
In the implementation presented here, the solenoidal constraint is maintained using constrained transport (CT).
These requirements become particularly demanding when simulations must also accommodate mesh refinement, heterogeneous computing architectures, additional microphysics, and, ultimately, dynamically evolving spacetimes \citep{Liska2022HAMR,Stone2024AthenaK}.

\begin{table*}[!t]
\caption{Representative framework capabilities and design choices, separated from the particular configurations compared numerically in this paper.
  Spacetime evolution distinguishes fixed from dynamical backgrounds; geometry access describes whether a prescribed metric is stored (static) or evaluated on access (dynamic), and coordinates/metrics lists the implemented coordinate or metric choices.
  These are distinct from evolving the Einstein equations.
  Entries refer to the cited implementations, rather than every later extension.
  \pangu{} I covers fixed backgrounds; the broader \pangu{} framework also includes dynamical spacetime, reserved for the companion paper.}
\label{tab:framework_comparison}
\centering
\footnotesize
\setlength{\tabcolsep}{2pt}
\renewcommand{\arraystretch}{1.3}
\noindent\makebox[\textwidth][l]{\hspace*{-0.04\textwidth}\resizebox{\textwidth}{!}{%
\begin{tabular}{@{}lccccccc@{}}
\toprule
Feature & \textbf{\pangu{} I} & \texttt{AthenaK} & \texttt{KHARMA} & \texttt{Phoebus} & \texttt{H-AMR} & \texttt{BHAC} & \texttt{GR-Athena++}\\
Reference & This work & \citep{Stone2024AthenaK} & \citep{Prather2024KHARMA} & \citep{Barker2024Phoebus} & \citep{Liska2022HAMR} & \citep{Porth2017BHAC} & \citep{Cook2023GRAthena}\\
\midrule
Spacetime evolution & fixed & fixed; dynamical & fixed & fixed; monopole gravity & fixed & fixed$^{a}$ & dynamical\\
Geometry access & static; dynamic & dynamic; NR$^{g}$ & static; dynamic & static; NR & static & static & NR\\
Coordinates/metrics & CKS; MKS; customized$^{b}$ & CKS; Minkowski & Minkowski; BL; MKS & analytic; tabulated & mapped KS & CKS; MKS; customized & Cartesian numerical metric\\
Foundation & \texttt{Parthenon}; \texttt{Kokkos} & \texttt{Kokkos} & \texttt{Parthenon}; \texttt{Kokkos} & \texttt{Parthenon}; \texttt{Kokkos} & \texttt{CUDA}/\allowbreak\texttt{OpenCL} & \texttt{MPI-AMRVAC} & \texttt{Athena++}\\
CT & Face CT & Face CT & Face CT; Flux CT & Cell-centered CT & Face CT & Face CT; Flux CT & Face CT\\
Refinement & SMR/AMR$^{c}$ & SMR/AMR & SMR/AMR & SMR/AMR & SMR/AMR & SMR/AMR & SMR/AMR\\
Electron physics & Multi-model heating & EOS; Compton exchange$^{d}$ & Multi-model heating & Composition; neutrinos & Two-temperature$^{e}$ & Heating extensions & No\\
MPI & Yes & Yes & Yes & Yes & Yes & Yes & Yes\\
GPU & Yes & Yes & Yes & Yes & Yes & No & No$^{f}$\\
\shortstack[l]{Portable\\Performance} & \shortstack{CPU/CUDA/\\HIP/SYCL} & \shortstack{CPU/CUDA/\\HIP/SYCL} & \shortstack{CPU/CUDA/\\HIP/SYCL} & \shortstack{CPU/CUDA/\\HIP/SYCL} & \shortstack{CUDA/\\OpenCL} & CPU only & CPU only\\
\bottomrule
\end{tabular}}}
\par\smallskip
\raggedright
$^{a}$\texttt{BHAC} denotes the stationary-spacetime implementation; extensions and parallelization are described separately \citep{Mizuno2021Electrons,BHAC2021Hybrid}.
$^{b}$Customized denotes a metric implementation supplied through the geometry interface; verification here covers CKS and MKS.
$^{c}$\pangu{} supports static/adaptive mesh refinement (SMR/AMR); the accretion tests here verify uniform meshes and CKS SMR. The current static-MKS path omits refinement.
$^{d}$Electron-fraction-dependent EOS and Compton energy exchange are distinct from a multi-model passive electron-heating module.
$^{e}$\texttt{H-AMR} couples electron/ion thermodynamics to radiation \citep{Liska2022TwoTemperature}.
$^{f}$The cited \texttt{GR-Athena++} implementation reports MPI/OpenMP CPU scaling; \texttt{AthenaK} GPU results are not attributed to \texttt{GR-Athena++}.
$^{g}$NR denotes numerical relativity.
\end{table*}

\begin{figure*}[t]
  \centering
  \includegraphics[width=0.93\textwidth]{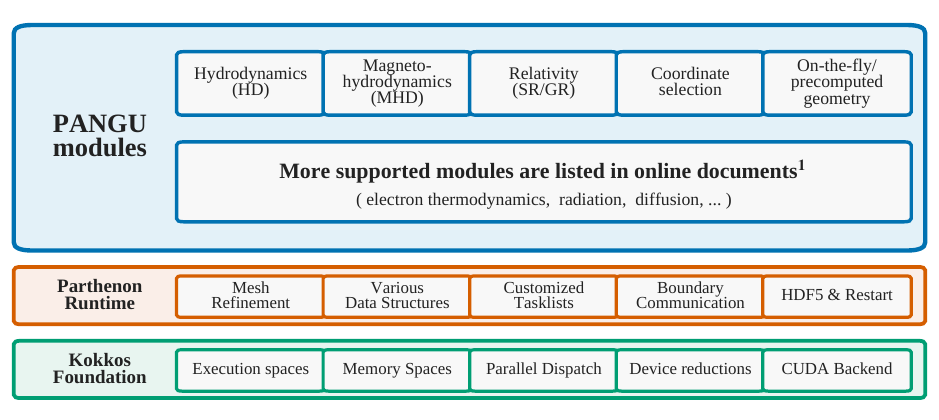}
  \caption{Software composition of \pangu{} I.
    Generic runtime services are supplied by \texttt{Parthenon} and \texttt{Kokkos}, while relativistic physics, coordinates, geometry access, and optional physics remain in \pangu{} packages.
    The supported modules are documented in the \pangu{} Wiki, \url{https://github.com/adamdarx/PANGU/wiki}.}
  \label{fig:architecture}
\end{figure*}

A broad ecosystem of relativistic-astrophysics codes has developed complementary capabilities rather than converging on a single software design.

Three development lineages are particularly relevant to the present work.
The first is the HARM lineage.
\texttt{HARM} introduced a compact conservative GRMHD formulation in modified Kerr--Schild coordinates \citep{Gammie2003HARM}, robust primitive-variable recovery followed \citep{Noble2006Primitive}, and \texttt{iharm3D} consolidated the scheme with modern verification and production workflows \citep{Prather2021Iharm3D}.
\texttt{H-AMR} carried this formulation to accelerator-resident accretion calculations \citep{Liska2022HAMR}, while \texttt{KHARMA} rebuilt it as a modular set of packages on \texttt{Parthenon} and \texttt{Kokkos}, with particular strength in mapped Kerr coordinates \citep{Prather2024KHARMA}.
The second is the Athena lineage.
\texttt{Athena} established an unsplit Godunov-type finite-volume scheme with face-centered CT \citep{Stone2008Athena}; \texttt{Athena++} rebuilt this design around block-structured static and adaptive mesh refinement and extended it to stationary-spacetime GRMHD \citep{Stone2020AthenaPP,White2016AthenaGRMHD}; \texttt{GR-Athena++} added numerical-relativity spacetimes \citep{Cook2023GRAthena}; and \texttt{AthenaK} reimplemented the framework with \texttt{Kokkos} as a purpose-built, performance-portable AMR code spanning fluid dynamics and numerical relativity \citep{Stone2024AthenaK}.
The third is the BHAC lineage.
Built on the \texttt{MPI-AMRVAC} toolkit, \texttt{BHAC} emphasizes GRMHD in general stationary metrics and coordinates, block-based mesh refinement, and independent CT formulations \citep{Porth2017BHAC}; later work extended it with hybrid parallelization \citep{BHAC2021Hybrid}.

\texttt{Phoebus}, \texttt{ECHO}, \texttt{Cosmos++}, \texttt{IllinoisGRMHD}, \texttt{KORAL}, and \texttt{PLUTO} provide further complementary formulations for stationary or dynamical spacetimes, radiation, composition, and relativistic flows \citep{Barker2024Phoebus,DelZanna2007ECHO,Anninos2005Cosmos,Etienne2015Illinois,Sadowski2013KORAL,Mignone2007PLUTO}.
Because these frameworks make different numerical and software choices, cross-code validation remains indispensable: agreement of a late turbulent snapshot is not equivalent to agreement of a deterministic discrete trajectory \citep{Porth2019CodeComparison}.
Table~\ref{tab:framework_comparison} summarizes these complementary design emphases and makes their software boundaries explicit without ranking physical fidelity or scientific scope.

The \texttt{Parthenon} framework \citep{Grete2023Parthenon}, which we credit with the generic infrastructure used throughout this work, provides the block-structured mesh, static and adaptive refinement, inter-block and MPI communication, task-graph execution, parameter input, I/O, and restart.
\texttt{Parthenon} generalized the block-structured mesh design of \texttt{Athena++} into a performance-portable foundation for downstream multiphysics applications, and \texttt{Kokkos} supplies the execution and memory abstractions that run the same kernels on CPUs and GPUs \citep{Edwards2014Kokkos,Trott2022Kokkos}.
Mesh refinement, parallel communication, and portable execution in this work therefore rest on these two frameworks.

\pangu{}, short for \textbf{P}arthenon-Based \textbf{A}strophysics for \textbf{N}umerical Relativity and \textbf{G}RMHD in a \textbf{U}nified Framework, is where these three lineages converge.
From the HARM lineage it adopts the conservative GRMHD formulation, modified Kerr--Schild coordinates, primitive recovery, and the \texttt{Parthenon} package model; from the Athena lineage, face-centered CT, the Cartesian Kerr--Schild and refined-mesh path, and a performance-portable design; and from the BHAC lineage, support for general stationary metrics behind a coordinate-independent geometry interface.
\pangu{} maintains an explicit boundary between infrastructure and astrophysical packages, as illustrated in Figure~\ref{fig:architecture}.
The contribution of \pangu{} I, the first paper of this series, is a single package architecture that combines a compact conservative formulation, portable block-structured refinement, mapped-coordinate geometry, and accelerator-resident execution.
It establishes the fixed-spacetime matter sector of the broader \pangu{} framework.
Numerical relativity, radiation transport, and the coupled Einstein--GRMHD system are outside the present scope and will be presented separately.

\pangu{} retains the astrophysical numerical layer in its own packages: reconstruction, Riemann solutions, face-centered CT, source terms, primitive recovery, repair policies, geometry handling, and optional physics.
This is a substantive software-design difference from an application that maintains its own complete AMR and runtime framework, while remaining complementary to the purpose-built, performance-portable AMR strategy demonstrated by \texttt{AthenaK} \citep{Stone2024AthenaK}.

Because the physics layer is shared, physical regime and spacetime representation do not define separate fluid solvers.
The same finite-volume and face-centered CT framework is used across Newtonian hydrodynamics and magnetohydrodynamics (HD and MHD), special-relativistic hydrodynamics (SRHD) and magnetohydrodynamics (SRMHD), and fixed-spacetime general-relativistic hydrodynamics (GRHD) and GRMHD.
Coordinate mapping, metric specification, and geometry-access policy are isolated from the fluid kernels and composed through a common location-aware interface, so geometry enters reconstruction, fluxes, source terms, primitive recovery, and diagnostics without requiring parallel implementations of the GRMHD algorithm.

\pangu{} I exercises this architecture in two deliberately different fixed-spacetime realizations.
Cartesian Kerr--Schild (CKS) geometry is evaluated at the point of use and combined with block-structured static mesh refinement, avoiding replication of stored geometry arrays across refined three-dimensional meshes.
Modified Kerr--Schild (MKS) geometry instead uses a symmetry-reduced stored representation that is advantageous for axisymmetric accretion calculations.
These configurations differ in coordinate representation, metric determinant, storage strategy, and intended computational regime, but expose the same geometry interface to the fluid solver.
They therefore demonstrate a practical compute--memory tradeoff within one GRMHD implementation rather than two independent coordinate-specific solvers.

Verification spans analytic convergence, Newtonian and relativistic waves and shocks, multidimensional MHD, and black-hole accretion, with comparisons to \texttt{AthenaK}, \texttt{KHARMA}, and \texttt{BHAC}.
Turbulent tests are assessed through morphology and integral diagnostics because independently seeded trajectories decorrelate.

Sections \ref{sec:physics} and \ref{sec:numerics} define the equations and their discretization.
Section~\ref{sec:verification} states the evidence protocol, and Secs.~\ref{sec:non_gr_results} and \ref{sec:gr_results} report verification from Newtonian waves through black-hole accretion.
The performance-baseline section reports a controlled single-GPU baseline and a separate A100 MPI strong-scaling campaign for a fixed global magnetized Bondi problem.
Limitations are stated in Sec.~\ref{sec:discussion}.
We use geometrized units $G=c=M=1$ and metric signature $(-,+,+,+)$ unless stated otherwise.
For black-hole tests, lengths and times are reported in $\rg\equiv GM/c^2$ and $\tg\equiv GM/c^3$, respectively; positions and times in Newtonian and special-relativistic tests are dimensionless.

\section{Physical model}
\label{sec:physics}

\subsection{Covariant ideal GRMHD}

Rest-mass and stress-energy conservation are
\begin{align}
 \nabla_\mu(\rho u^\mu)&=0,                                      \label{eq:mass}\\
 \nabla_\mu T^{\mu}{}_{\nu}&=0,                                 \label{eq:stresscons}
\end{align}
where $\rho$ is the rest-mass density and $u^\mu$ is the contravariant fluid four-velocity, normalized by $u^\mu u_\mu=-1$.
Maxwell's homogeneous equation and the ideal-conductivity condition are
\begin{align}
 \nabla_\mu {{}^*F}^{\mu\nu}&=0,                                \label{eq:maxwell}\\
 F^{\mu\nu}u_\nu&=0,                                            \label{eq:ideal}
\end{align}
where $F^{\mu\nu}$ and ${}^*F^{\mu\nu}$ are the Faraday tensor and its dual.
The magnetic four-vector measured in the fluid frame is $b^\mu=-{}^*F^{\mu\nu}u_\nu$, and $b^2=b^\mu b_\mu$.
For an ideal gas,
\begin{equation}
 p=(\Gamma-1)u_{\rm g},\qquad
 h=1+\frac{u_{\rm g}+p}{\rho},                                  \label{eq:eos}
\end{equation}
where $u_{\rm g}$ is the internal-energy density, $p$ is gas pressure, $\Gamma$ is the adiabatic index, and $h$ is specific enthalpy.
The ideal-GRMHD stress tensor is
\begin{equation}
 T^{\mu\nu}=(\rho h+b^2)u^\mu u^\nu
 +\left(p+\frac{b^2}{2}\right)g^{\mu\nu}-b^\mu b^\nu .          \label{eq:stress}
\end{equation}
The magnetic pressure is $b^2/2$.
Equations~\eqref{eq:mass}--\eqref{eq:stress} reduce continuously to relativistic hydrodynamics when $b^\mu=0$, to SRMHD on a Minkowski metric, and to the familiar Newtonian HD/MHD systems when both field and flow speeds are nonrelativistic.
These limits share one code-level state convention rather than separate physical interpretations.

\subsection{Coordinate conservative formulation}

For numerical calculation, choose a coordinate chart $x^\mu=(t,x^i)$, let $g\equiv\det(g_{\mu\nu})$, and define the coordinate magnetic field $B^i\equiv{}^*F^{i0}$. Its relation to the fluid-frame magnetic four-vector is
\begin{equation}
 b^0=u_iB^i,\qquad b^i=\frac{B^i+b^0u^i}{u^0}.             \label{eq:b4}
\end{equation}
Using these relations, the covariant equations are expanded in the chosen chart and written as the single hyperbolic balance law
\begin{equation}
 \partial_t\bm U+\partial_i\bm F^i=\bm S, \label{eq:master_balance}
\end{equation}

For compactness, define
\begin{equation}
 \begin{aligned}
 D^\mu&\equiv\rho u^\mu,
 &\mathcal T^\mu{}_\nu&\equiv T^\mu{}_\nu+D^\mu\delta^0{}_\nu,\\
 \mathcal F_B^{ij}&\equiv2b^{[j}u^{i]},
 &\mathcal G_\nu&\equiv
 \frac{1}{2}T^{\kappa\lambda}\partial_\nu g_{\kappa\lambda},
 \end{aligned}
                                                                    \label{eq:transport_blocks}
\end{equation}
Here $\delta^\mu{}_\nu$ is the Kronecker delta, so $\delta^0{}_\nu$ is unity for $\nu=0$ and zero otherwise.
Antisymmetrization includes the factor $1/2$, such that $2b^{[j}u^{i]}=b^ju^i-b^iu^j$.
With these definitions, the conserved state, flux, and source for a prescribed stationary metric are
\begin{align}
 \bm U&=\sqrt{-g}\left(D^0,\mathcal T^0{}_1,\mathcal T^0{}_2,\mathcal T^0{}_3,\mathcal T^0{}_0,B^1,B^2,B^3\right)^{\mathsf T}, \label{eq:state}\\
 \bm F^i&=\sqrt{-g}\left(D^i,\mathcal T^i{}_1,\mathcal T^i{}_2,\mathcal T^i{}_3,\mathcal T^i{}_0,\mathcal F_B^{i1},\mathcal F_B^{i2},\mathcal F_B^{i3}\right)^{\mathsf T}, \label{eq:grflux}\\
 \bm S&=\sqrt{-g}\left(0,\mathcal G_1,\mathcal G_2,\mathcal G_3,\mathcal G_0,0,0,0\right)^{\mathsf T}. \label{eq:source_vector}
\end{align}
Here $\bm U$ is the densitized conserved state, $\bm F^i$ is the continuous physical flux density in direction $i$, and $\bm S$ is the geometric source.
For a stationary metric, $\partial_0g_{\kappa\lambda}=0$ and hence $\mathcal G_0=0$, so $\bm S$ contains only the three spatial momentum sources.
The fifth state component, $\mathcal T^0{}_0$, carries a covariant time index and therefore equals the negative of the usual rest-mass-subtracted total-energy density in Minkowski spacetime.
Antisymmetry gives $\mathcal F_B^{ii}=0$, while $\mathcal T^\mu{}_\nu$ and $\mathcal G_\nu$ serve only as shorthand; the physical stress-energy tensor remains $T^{\mu\nu}$.

Equation~\eqref{eq:master_balance} advances the conserved state $\bm U$, whereas the fluxes $\bm F^i$ and source $\bm S$ cannot in general be evaluated explicitly from $\bm U$.
All three can instead be constructed from the primitive-variable vector
\begin{equation}
 \bm P=\left(\rho,\,u_{\rm g},\,\widetilde u^1,\widetilde u^2,
 \widetilde u^3,\,B^1,B^2,B^3\right)^{\mathsf T}.              \label{eq:primitive_state}
\end{equation}
Following \citet{McKinneyGammie2004}, we use the Eulerian-frame spatial four-velocity $\widetilde u^i$.
Its relation to the coordinate four-velocity in the 3+1 decomposition \citep{Porth2017BHAC} is
\begin{equation}
 u^0=\frac{W}{\alpha},\qquad
 u^i=\widetilde u^i-\frac{W}{\alpha}\beta^i,                  \label{eq:u4}
\end{equation}
where $W=\sqrt{1+\gamma_{ij}\widetilde u^i\widetilde u^j}$ is the Lorentz factor, $\alpha=(-g^{00})^{-1/2}$ is the lapse function, $\beta^i=\alpha^2g^{0i}$ is the shift vector, and $\gamma_{ij}=g_{ij}$ is the spatial metric.
Together with Eqs.~\eqref{eq:eos} and \eqref{eq:b4}, $\bm P$ determines every component in Eqs.~\eqref{eq:state}--\eqref{eq:source_vector}.
Pressure follows Eq.~\eqref{eq:eos} and is not an additional independent primitive variable.
The 3+1 construction and the numerical reason for using $\widetilde u^i$ instead of $u^i$ as primitive variables are given in Appendix~\ref{app:primitive_velocity}.

\section{Numerical methods}
\label{sec:numerics}

Equation~\eqref{eq:master_balance} gives the continuous balance law. \pangu{} solves it with a finite-volume discretization: coordinate space is partitioned into control volumes, the spatial divergence is replaced by numerical fluxes across their faces, and the resulting cell averages are advanced in time.

\subsection{Finite-volume update}

Let $i$ be the multi-index of a computational cell with coordinate volume $V_i$, and let $f\in\partial i$ index the faces on its boundary, each with coordinate-face measure $A_f$. For a time step $\Delta t$, $\bm U_i^{(n)}$ denotes the cell-averaged conserved state at Runge--Kutta stage $n$, and $\bm S_i$ is the cell-centered source. The Riemann solver assigns a single outward-oriented numerical flux $\widehat{\bm F}_f$ to each face. With these spatially discrete quantities, the conservative update is
\begin{equation}
 \bm U_i^{(n+1)}=\bm U_i^{(n)}-
 \frac{\Delta t}{V_i}\sum_{f\in\partial i}A_f\widehat{\bm F}_f
 +\Delta t\,\bm S_i,                                             \label{eq:fv}
\end{equation}

\begin{figure*}[!t]
  \centering
  \begin{minipage}[t]{0.98\textwidth}\centering
    {\large\textbf{(a)}}\\[0.6ex]
    \includegraphics[width=0.96\linewidth]{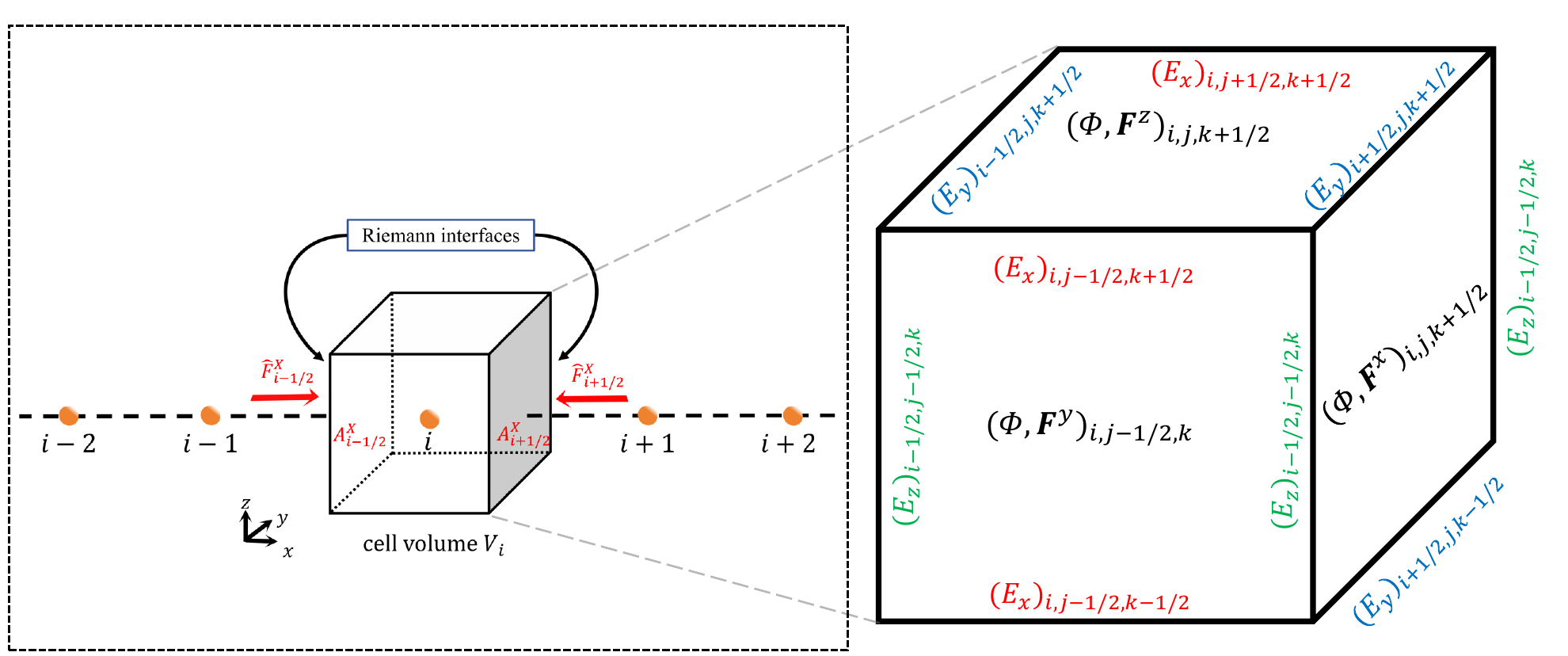}
  \end{minipage}
  \par\vspace{1.0ex}
  \begin{minipage}[t]{0.98\textwidth}\centering
    {\large\textbf{(b)}}\\[0.6ex]
    \figframe{\includegraphics[width=0.97\linewidth]{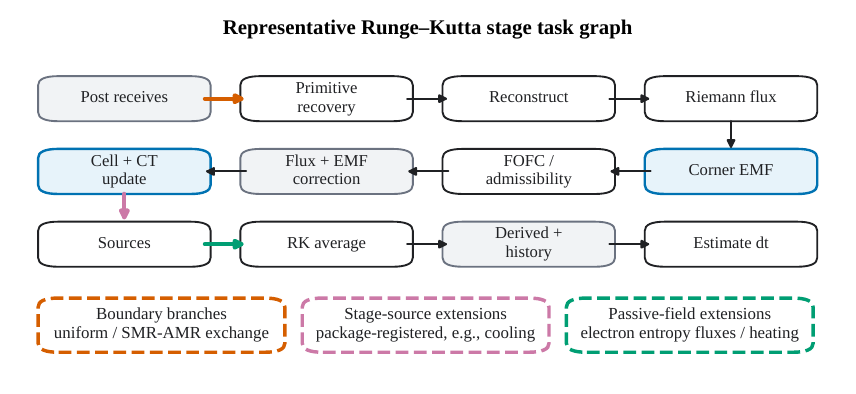}}
  \end{minipage}
  \caption{Discretization and task organization.
    (a) Finite-volume control volume, shared face fluxes, and face-centered CT staggering, adapted from the \texttt{BHAC} description and accompanying online schematic \citep{Porth2017BHAC}.
    (b) Data dependencies within one Runge--Kutta stage; package tasks operate on packed \texttt{MeshData} across blocks.}
  \label{fig:fv_cell}\label{fig:ct_staggering}\label{fig:tasks}
\end{figure*}

The cell-and-face bookkeeping is illustrated in Figure~\ref{fig:fv_cell}(a).
The code uses a two-stage second-order Runge--Kutta method (RK2) in the tests reported here \citep{Godunov1959,ShuOsher1988RK}.
Equation~\eqref{eq:fv} is applied dimension by dimension using one numerical flux per face, so adjacent cells receive equal and opposite flux contributions and the finite-volume update remains conservative.

Reconstruction is necessary because Eq.~\eqref{eq:fv} evolves cell averages, whereas an upwind Riemann solver requires distinct left and right limiting states at each face.
\pangu{} reconstructs primitive rather than conserved components, keeping density and internal energy directly available to the limiter and avoiding a nonlinear conservative-to-primitive inversion of an interpolated state.
The normal face magnetic field is supplied by CT; the remaining fluid and magnetic components are reconstructed from cell centers.
The face metric is sampled after the two states have been formed.

For a scalar cell average $q_i$, the default piecewise-linear method (PLM) uses the harmonic (van Leer) interface increment defined below.
With $\Delta_-=q_i-q_{i-1}$ and $\Delta_+=q_{i+1}-q_i$,
\begin{equation}
 \delta q_i=
 \begin{cases}
 \displaystyle\frac{\Delta_-\Delta_+}{\Delta_-+\Delta_+},
 &\Delta_-\Delta_+>0,\\[5pt]
 0,&\text{otherwise},
 \end{cases}                                                       \label{eq:plm}
\end{equation}
with face values $q_{i+1/2,L}=q_i+\delta q_i$ and $q_{i+1/2,R}=q_{i+1}-\delta q_{i+1}$.
 Donor-cell, the fourth-order-interface piecewise parabolic method (PPM4), extremum-preserving PPM, and weighted essentially non-oscillatory WENO-Z paths are also compiled.
The PPM path first forms
\begin{equation}
 q_{i+1/2}^{(4)}=\frac{7}{12}(q_i+q_{i+1})
 -\frac{1}{12}(q_{i-1}+q_{i+2}),                                  \label{eq:ppm4}
\end{equation}
clips this interface value to the interval spanned by its two cells, and then limits each cell parabola.
Four ghost zones are required for relativistic PPM; near a physical boundary the ghost fill precedes reconstruction, so the same stencil is used rather than an undocumented order reduction \citep{VanLeer1979MUSCL,ColellaWoodward1984PPM}.

\subsection{Riemann solver}

At each face, \pangu{} converts the reconstructed left and right physical states into one common upwind flux by solving a local one-dimensional Riemann problem.
For relativistic HD and MHD, the available approximate solvers are the local Lax--Friedrichs solver (LLF; also LAXF or Rusanov) and the less diffusive two-wave Harten--Lax--van Leer--Einfeldt (HLLE) solver \citep{Lax1954,Rusanov1962,Harten1983HLL,Einfeldt1988HLLE}.
Let $\bm U_L$ and $\bm U_R$ denote the conserved states obtained from the reconstructed left and right states, and let $\bm F_L$ and $\bm F_R$ be their corresponding physical fluxes. The quantities $\lambda_{L,R}^-$ and $\lambda_{L,R}^+$ are the left- and right-going characteristic speeds evaluated in those states.
LLF uses one symmetric signal bound $a=\max(|\lambda_L^-|,|\lambda_L^+|, |\lambda_R^-|,|\lambda_R^+|)$,
\begin{equation}
 \widehat{\bm F}_{\rm LLF}=\frac{\bm F_L+\bm F_R}{2}
 -\frac{a}{2}(\bm U_R-\bm U_L),                                  \label{eq:llf}
\end{equation}
 and is also the deliberately diffusive fallback used by the first-order flux-correction procedure (FOFC; Sec.~\ref{sec:fofc}).
 The default relativistic production choice is HLLE. With $\lambda_-\le0\le\lambda_+$ denoting lower and upper bounds on the characteristic speeds from the two reconstructed states, its numerical flux is
\begin{equation}
 \widehat{\bm F}_{\rm HLLE}=
 \frac{\lambda_+\bm F_L-\lambda_-\bm F_R
 +\lambda_+\lambda_-(\bm U_R-\bm U_L)}{\lambda_+-\lambda_-},      \label{eq:hlle}
\end{equation}

For GRMHD, let $c_s$ and $c_A$ denote the sound and Alfv\'en speeds, respectively. \pangu{} forms the approximate fast-magnetosonic speed $c_f^2=c_s^2+c_A^2-c_s^2c_A^2$, inserts it into the metric-dependent characteristic quadratic, and bounds the result by the coordinate light cone \citep{Gammie2003HARM,Komissarov1999RMHD}.

The Newtonian packages expose a broader solver set because no relativistic primitive inversion is involved: HD provides LLF, HLLE, HLLC, and Roe, while MHD provides LLF, HLLE, and HLLD.
The relativistic verification in this paper uses HLLE except where a test or FOFC explicitly selects LLF.

\subsection{Time-step control}

The flux solver couples adjacent states, whereas the time-step estimator independently uses signal speeds to enforce the Courant--Friedrichs--Lewy (CFL) stability limit.
\pangu{} provides two time-step estimates, denoted \texttt{light} and \texttt{wave}.

\paragraph{Newtonian coordinates.}

For computational cell $i$ and active direction $d\in\{1,\ldots,N_{\rm dim}\}$, where $N_{\rm dim}$ is the number of active dimensions, let $\Delta x_{i,d}$ be the cell width, $v_{i,d}$ the fluid velocity, and $c_{i,d}$ the sound speed for HD or directional fast speed for MHD. With prescribed Courant number $C_{\rm CFL}$, the directional bound is
\begin{equation}
 \Delta t_{\rm N}=C_{\rm CFL}\min_{i,d}
 \frac{\Delta x_{i,d}}{|v_{i,d}|+c_{i,d}},                       \label{eq:cflnewtonian}
\end{equation}

An enabled diffusion operator can impose a smaller parabolic step.

\paragraph{Relativistic estimates.}

For a relativistic package, let $a_{i,d}=\max(|\lambda_{d,+}|,|\lambda_{d,-}|)$ be the coordinate signal-speed bound used in cell $i$ and direction $d$. The \texttt{light} path applies the directional minimum $\Delta t=C_{\rm CFL}\min_{i,d}(\Delta x_{i,d}/a_{i,d})$.

Suppressing the cell index inside the minimum, the \texttt{wave} path uses the unsplit multidimensional reciprocal sum
\begin{equation}
 \Delta t=C_{\rm CFL}\min_{\rm cells}
 \left[\sum_{d=1}^{N_{\rm dim}}
 \frac{a_d}{\Delta x_d}\right]^{-1}.                            \label{eq:cflwave}
\end{equation}
The directional minimum is cheaper and less restrictive in multiple dimensions; the reciprocal sum accounts for simultaneous propagation along all active axes and is consequently more conservative.
Although the CFL condition retains this common algebraic form, the admissible step depends on coordinate signal speeds set by the metric and coordinate chart, together with coordinate cell widths set by the grid mapping. Accordingly, \pangu{} uses metric-independent relativistic characteristics in Cartesian Minkowski coordinates, permits either a unit-coordinate-light-speed or wave-speed estimate in CKS coordinates, and requires metric-dependent wave speeds in MKS coordinates. The coordinate-specific considerations are summarized in Appendix~\ref{sec:cfl_coordinates}.

\subsection{Geometry composition}

Metric and access mode are independent compile-time choices.
The metric class supplies the maps and tensors defined in Sec.~\ref{sec:metrics}; the mode says how a kernel receives them.
\emph{dynamic} mode evaluates a prescribed metric at the point of use, whereas \emph{static} mode materializes the metric, inverse, determinant, lapse, shift, and derivatives at cell and face locations.
Both expose the same location-aware interface, including determinants collocated with the flux or state they densitize.
In this paper, three-dimensional CKS uses dynamic access with static mesh refinement (SMR) and excision, whereas axisymmetric MKS uses static access without refinement to limit metric storage.
Both operate on prescribed stationary backgrounds; synchronized Einstein--matter evolution is reserved for Paper II.

\subsection{Primitive recovery and admissibility}
\label{sec:c2p}

Godunov fluxes evolve the conserved state, but reconstruction, the equation of state, characteristic speeds, and diagnostics require primitive variables.
In relativistic MHD the map cannot be inverted component by component: the unknown Lorentz factor and enthalpy enter the momentum and energy together, and the magnetic terms couple them nonlinearly.
Recovery is therefore an essential nonlinear solve after every Runge--Kutta update.
It is also a principal robustness risk because truncation error can make low-density or highly magnetized conserved states temporarily inadmissible.

\pangu{} uses the robust one-dimensional construction of Kastaun et al.~\citep{Kastaun2021Primitive}.
After undensitizing the GR state and projecting it into the normal frame, scalar contractions of the momentum and magnetic field reduce the coupled inversion to one bounded unknown $0\le\mu\le1$.
Solving that residual determines the Lorentz factor and hence the density, internal energy, pressure, and spatial four-velocity.
The residual, admissible bracket, and uniqueness argument are given in Ref.~\citep{Kastaun2021Primitive}; we do not repeat their algebra here.
\pangu{} solves the bracketed problem with the Illinois false-position method, a $10^{-12}$ tolerance, and at most 25 iterations.
 A result is accepted only if the reconstructed state satisfies the equation of state (EOS) and physical bounds; otherwise the failure is recorded for repair.
This bounded formulation avoids a multidimensional initial guess and complements other GRMHD inversion schemes \citep{Noble2006Primitive,Siegel2018Primitive}.

Admissibility is hierarchical.
 Before inversion, density and energy are raised to configured floors; the effective density floor can enforce the configured magnetization ceiling $\sigma_{\max}$ through $\rho\ge b^2/\sigma_{\max}$.
After inversion, pressure/entropy and Lorentz-factor ceilings are applied.
A successful repair is immediately converted back to conserved form, so the next stage does not carry an inconsistent primitive--conserved pair.

\subsection{First-order flux correction}
\label{sec:fofc}

First-order flux correction (FOFC) is applied before the final recovery.
A candidate high-order update is inverted cell by cell.
Cells that fail or activate a floor, Lorentz ceiling, magnetization ceiling, or excision guard are flagged.
Every face adjacent to a flagged cell has both its fluid flux and its face electromotive force (EMF) replaced by the donor-cell/LLF counterpart.
The update and recovery are then repeated.
Thus FOFC is a local conservative replacement of the flux divergence and CT curl, not a post-update smoothing operation.

\subsection{Face-centered CT}

In the continuum induction equation, an initially solenoidal magnetic field remains solenoidal because the divergence of a curl vanishes.
A floating-point finite-volume calculation does not inherit this identity automatically: reconstruction, independently evaluated face fluxes, roundoff, and mesh-level transfers can create a nonzero discrete divergence even when the analytic solution satisfies $\nabla\!\cdot\!\bm B=0$.
Such numerical monopoles can exert spurious magnetic forces and contaminate shocks and magnetically dominated accretion flows \citep{BrackbillBarnes1980,Toth2000DivB}.

\pangu{} uses face-centered CT for this purpose.
It evolves the magnetic flux through each face with edge-centered EMFs shared by all incident faces, so the cancellation follows from mesh topology rather than from separately approximated derivatives.
Magnetic evolution is an optional package: HD omits the three magnetic arrays and CT operations, whereas MHD and GRMHD retain the complete magnetic subsystem.
For a face $f$ normal to coordinate direction $i(f)$, let $\mathcal B_f$ be the face average of $\sqrt{-g}B^{i(f)}$, and denote the collection of these face fields by $\bm{\mathcal B}$. The corresponding oriented magnetic flux is $\Phi_f\equiv A_f\mathcal B_f$.
 Let $e\in\partial f$ index the oriented boundary edges of face $f$, and let $\mathcal E_e$ and $\Delta l_e$ denote the edge-centered EMF and edge length. The CT update is
\begin{equation}
 \frac{\dd\Phi_f}{\dd t}=-\sum_{e\in\partial f}\mathcal E_e\Delta l_e,
                                                                    \label{eq:ct}
\end{equation}
 where the same edge EMF is shared by every face incident on that edge \citep{Evans1988CT,GardinerStone2005,BalsaraSpicer1999CT,Toth2000DivB, DelZanna2003UCT,LondrilloDelZanna2004UCT,FelkerStone2018UCT}.
 Because the discrete boundary of a boundary vanishes, let $s_f=\pm1$ encode the outward orientation of face $f$ relative to cell $i$. The signed face divergence
\begin{equation}
 (\nabla\!\cdot\!\bm{\mathcal B})_i=
 \frac{1}{V_i}\sum_{f\in\partial i}s_f\mathcal B_fA_f             \label{eq:divb}
\end{equation}
is preserved to roundoff.
\pangu{} constructs corner EMFs from the directional face fluxes, exchanges the edge field across block boundaries, and only then updates the face field.
This ordering prevents adjacent blocks from using two values for the same topological edge.

\subsection{SMR, task ordering, and restart}

On SMR meshes, cell-centered conserved quantities use conservative averaging under restriction and slope-limited shared prolongation.
Face magnetic fields use area-consistent restriction, shared-face prolongation, and the internal T\'oth--Roe divergence-preserving construction \citep{TothRoe2002}.
The flux-correction stage in the \texttt{Parthenon} framework synchronizes coarse and fine fluxes before the cell and face updates.
The validation scope in this paper is restricted to uniform and statically refined meshes.

Figure~\ref{fig:tasks}(b) locates these operations within one Runge--Kutta stage. Reconstruction and Riemann solutions supply the candidate high-order fluxes, while CT edge synchronization and coarse--fine flux correction precede the candidate conservative update. Provisional recovery, FOFC (Sec.~\ref{sec:fofc}), physical-boundary application, and final recovery then follow in sequence.
The optional passive electron-entropy module is registered only when selected; its implementation sequence and regression tests are documented in the \pangu{} Wiki, \url{https://github.com/adamdarx/PANGU/wiki}.
Restart files contain the independent cell and face fields, and derived states are rebuilt after restart.

\section{Verification methodology}
\label{sec:verification}

Table~\ref{tab:verification} maps each numerical regime to its comparator and acceptance evidence; the paragraphs below define how those criteria are applied.

\begin{table*}[t]
\caption{Verification matrix. ``Matched'' means the discrete choices and stored times are identical; ``native'' retains each comparison code's own documented operator.
  SMR denotes static mesh refinement. CPAW denotes circularly polarized Alfv\'en wave.}
\label{tab:verification}
\begin{ruledtabular}
\scriptsize
\begin{tabular*}{\textwidth}{@{\extracolsep{\fill}}lllll}
Regime & Representative problem & Geometry/mesh & Comparator & Acceptance evidence\\
\hline
HD & smooth waves, Sod, blast & Cartesian uniform & analytic, \texttt{AthenaK} & order, trajectory, symmetry\\
MHD & CPAW, Orszag--Tang & Cartesian uniform & analytic, \texttt{AthenaK} & order, $\nabla\!\cdot\!\bm{\mathcal B}$, morphology\\
SRHD/SRMHD & eigenwaves, MUB1, blast & Minkowski uniform & analytic, \texttt{AthenaK}, \texttt{KHARMA} & order, matched/native profiles\\
GRHD & Bondi & CKS+SMR; MKS uniform & analytic, \texttt{AthenaK}, \texttt{KHARMA} & steady drift, order, trajectory\\
GRMHD & magnetized Bondi, monopole & CKS+SMR; MKS uniform & analytic, \texttt{AthenaK}, \texttt{KHARMA} & $\sigma$, flux, horizon diagnostics\\
GRMHD & SANE torus & CKS+SMR; MKS uniform & \texttt{AthenaK}, \texttt{BHAC} data & morphology, $\dot M$, $\Phi_B$\\
 &  &  &  & \\
\end{tabular*}
\end{ruledtabular}
\end{table*}

For smooth analytic problems we report $L_1$ errors and adjacent-resolution orders.
For discontinuities we report wave ordering, plateau states, and localized norms, because pointwise convergence at a moving shock is not expected.
Matched cross-code comparisons use identical mesh coordinates, initial values, reconstruction, Riemann solver, Runge--Kutta coefficients, CFL number, and output events.
Each saved event is compared; a final-time overlap alone is insufficient.
Native cross-code comparisons retain documented algorithmic differences and are interpreted through convergence and physical diagnostics.

 Reported differences are absolute double-precision (FP64) differences of the named primitive variable unless labeled relative.
A matched $L_\infty$ difference is the maximum over all compared cells and stored events, and the quoted $L_1$ difference is the largest per-event cell mean.
The scale of a single floating-point roundoff error is the IEEE-754 double-precision machine epsilon, $\epsilon=2.22\times10^{-16}$.
Matched trajectories accumulate such errors over many cycles and cells, so their maximum differences can exceed $\epsilon$ by one to two orders of magnitude.
We therefore quote the measured difference together with the cycle count, and attribute it to accumulated roundoff only when it remains many orders of magnitude below the discretization error of the same calculation.
The face-centered CT divergence, Eq.~\eqref{eq:divb}, is stored without normalization. For the Cartesian tests below, where $\sqrt{-g}=1$, a dimensionless value is quoted as $\max|\nabla\!\cdot\!\bm{\mathcal B}|\,\Delta x/\max|\bm B|$.

 For turbulent disks, stochastic perturbations and nonlinear growth of the magnetorotational instability (MRI) make late-time pointwise agreement inappropriate.
We instead report rest-mass accretion rate, magnetic flux, and morphology.
For the event horizon $\mathcal H$, let $(t,\xi^\perp,\eta^1,\eta^2)$ denote a surface-adapted diagnostic chart, where $\xi^\perp$ increases along the outward surface normal and $(\eta^1,\eta^2)$ parametrize the horizon. All components and the metric measure below are evaluated in this chart. We define the accretion rate and unsigned magnetic flux by
\begin{equation}
 \begin{aligned}
 \dot M &=-\int_{\mathcal H}\rho u^\perp\sqrt{-g}\,\dd\eta^1\dd\eta^2,\\
 \Phi_B &=\frac{1}{2}\int_{\mathcal H}|B^\perp|\sqrt{-g}\,\dd\eta^1\dd\eta^2 .
 \end{aligned}
 \label{eq:sane_diag}
\end{equation}
The sign convention makes $\dot M>0$ for inward flow with $u^\perp<0$, and the factor $1/2$ gives an unsigned hemispheric flux.
Unless explicitly defined as dimensionless, $\Phi_B$ below is the unsigned hemispheric magnetic flux in the simulation normalization.
A standard and normal evolution (SANE)/magnetically arrested disk (MAD) classification would require the dimensionless quantity $\phi=\Phi_B/\sqrt{|\dot M|}$ and a statistically stationary interval \citep{Narayan2012SANE,Tchekhovskoy2011MAD}; no such claim is inferred from the early-time curves alone.

Electron-module implementation and regression tests are documented in the \pangu{} Wiki; they lie outside the validation evidence presented in this paper.

Every figure is generated from archived output by a separate analysis script.
The MKS three-dimensional horizon reductions are additionally stored as CSV, so regenerating typography does not repeat the surface integral.
Input, executable, mesh, output index, and physical-time metadata are retained with the reductions.
Results labeled CKS exclude the excision/shear layer unless the layer itself is the subject of the test.

\section{Newtonian and special-relativistic verification}
\label{sec:non_gr_results}

\subsection{HD and MHD}

\begin{figure}[!b]
  \centering
  \figframe{\includegraphics[width=0.96\columnwidth]{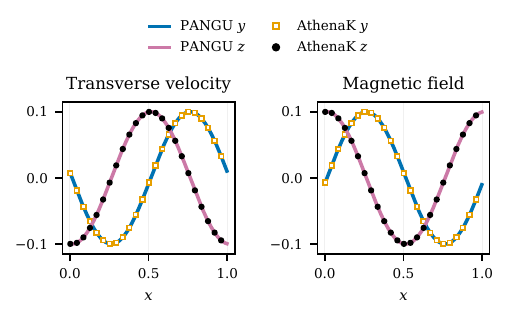}}
  \caption{Circularly polarized Alfv\'en wave (CPAW).
    The left panel compares the transverse velocity components $v_y$ and $v_z$ from \pangu{} and \texttt{AthenaK}; the right panel compares the corresponding magnetic-field components $B_y$ and $B_z$.
    \pangu{} uses blue and pink curves for the $y$ and $z$ components, respectively, while \texttt{AthenaK} uses square and circular markers for the same components.
    Profiles are the final archived samples at $t=0.014$ on $192\times48\times48$ meshes; every eighth \texttt{AthenaK} sample is marked.
    Positions and times are dimensionless.
    The matched CT calculations agree closely while preserving the face-field divergence constraint.}
  \label{fig:cpaw}
\end{figure}

\begin{figure*}[!t]
  \centering
  \figframe{\includegraphics[width=0.96\textwidth]{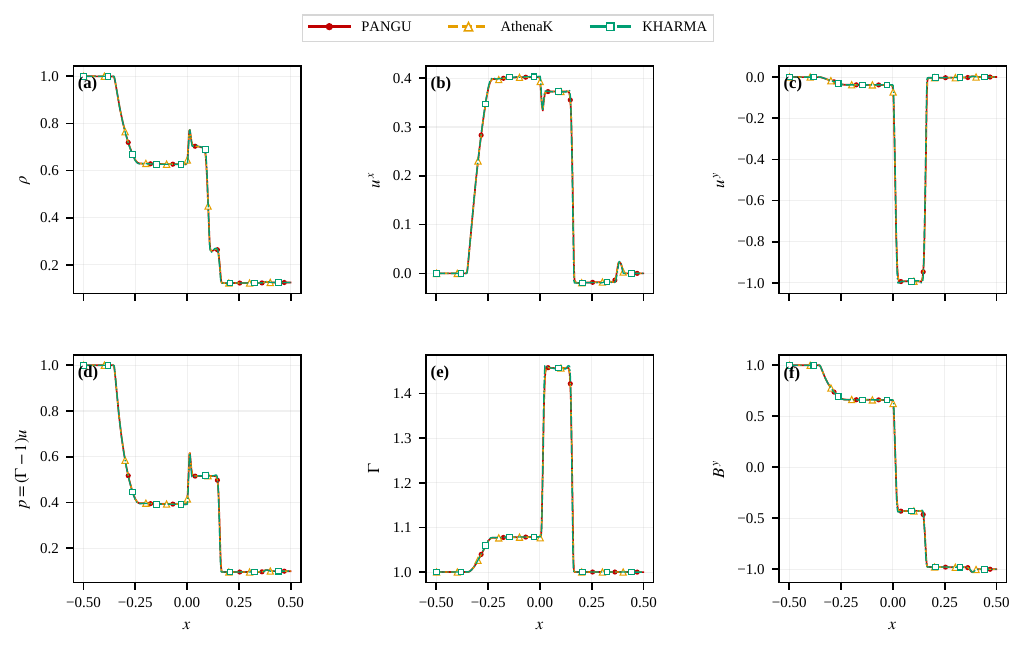}}
  \caption{MUB1 profiles at $t=0.4$.
    The matched \pangu{}--\texttt{AthenaK} profiles differ by at most $1.33\times10^{-14}$, while the \texttt{KHARMA} differences produced by its native reconstruction and CT path remain localized at moving discontinuities.
    The plotted velocity is the spatial four-velocity.
    Positions and times are dimensionless.}
  \label{fig:srmhd}
\end{figure*}

The Newtonian tests separate smooth-order verification, one-dimensional shock trajectories, and multidimensional robustness; Table~\ref{tab:verification} identifies the comparator and acceptance evidence for each group.
Smooth acoustic and entropy waves are evolved for one period with PLM, HLLE, and RK2 at $N=64$--512.
The entropy-wave density $L_1$ error decreases from $9.31\times10^{-4}$ to $1.42\times10^{-5}$, with adjacent orders 1.96, 2.04, and 2.03.
With the default acoustic amplitude $A=10^{-4}$, the orders are 1.96 and 1.90 but fall to 1.04 between $N=256$ and 512, where the error ($3.25\times10^{-8}$) is comparable to $A^2$, consistent with the nonlinear error of the finite-amplitude eigenmode.
Reducing the amplitude to $A=10^{-6}$ gives orders 1.96, 2.01, and 2.00.

The matched Sod problem \citep{Sod1978} uses 512 cells in four MeshBlocks, donor-cell reconstruction, HLLE, RK2, and CFL 0.4, and is compared at five events through $t=0.2$ (556 cycles).
The largest \pangu{}--\texttt{AthenaK} pointwise difference is $6.72\times10^{-15}$ (velocity, $\approx30\epsilon$), and the largest $L_1$ difference is $4.09\times10^{-16}$.
The two codes have the same final density $L_1$ error against the exact solution, $7.50\times10^{-3}$, to within $4\times10^{-17}$, so the inter-code difference lies twelve orders of magnitude below the discretization error.
Multidimensional blast, Kelvin--Helmholtz, and Rayleigh--Taylor calculations remain finite and preserve the expected symmetries, but are not used to infer smooth-solution order.

The circularly polarized Alfv\'en wave (CPAW) is a stringent matched test: the transverse velocity and magnetic components must retain a common phase while CT evolves all three face orientations.
Figure~\ref{fig:cpaw} shows the overlaid transverse profiles and the corresponding trajectory differences.
The matched calculation uses one $192\times48\times48$ MeshBlock, PLM, HLLE, RK2, and CFL 0.3, and is compared at five events through $t=0.014$ (10 cycles).
Among all events and all three velocity components, the three cell-centered field components, density, and pressure, the largest pointwise difference is $2.00\times10^{-14}$ (in $B_y$, $\approx90\epsilon$) and the largest $L_1$ difference is $5.86\times10^{-15}$.
The \pangu{} face-CT divergence is exactly zero at every stored event.
The close agreement and roundoff-level face-field divergence demonstrate both the propagation accuracy and the preservation of the discrete magnetic constraint.

\subsection{SRHD and SRMHD}

\begin{figure*}[t]
 \centering
 \figframe{\includegraphics[width=0.96\textwidth]{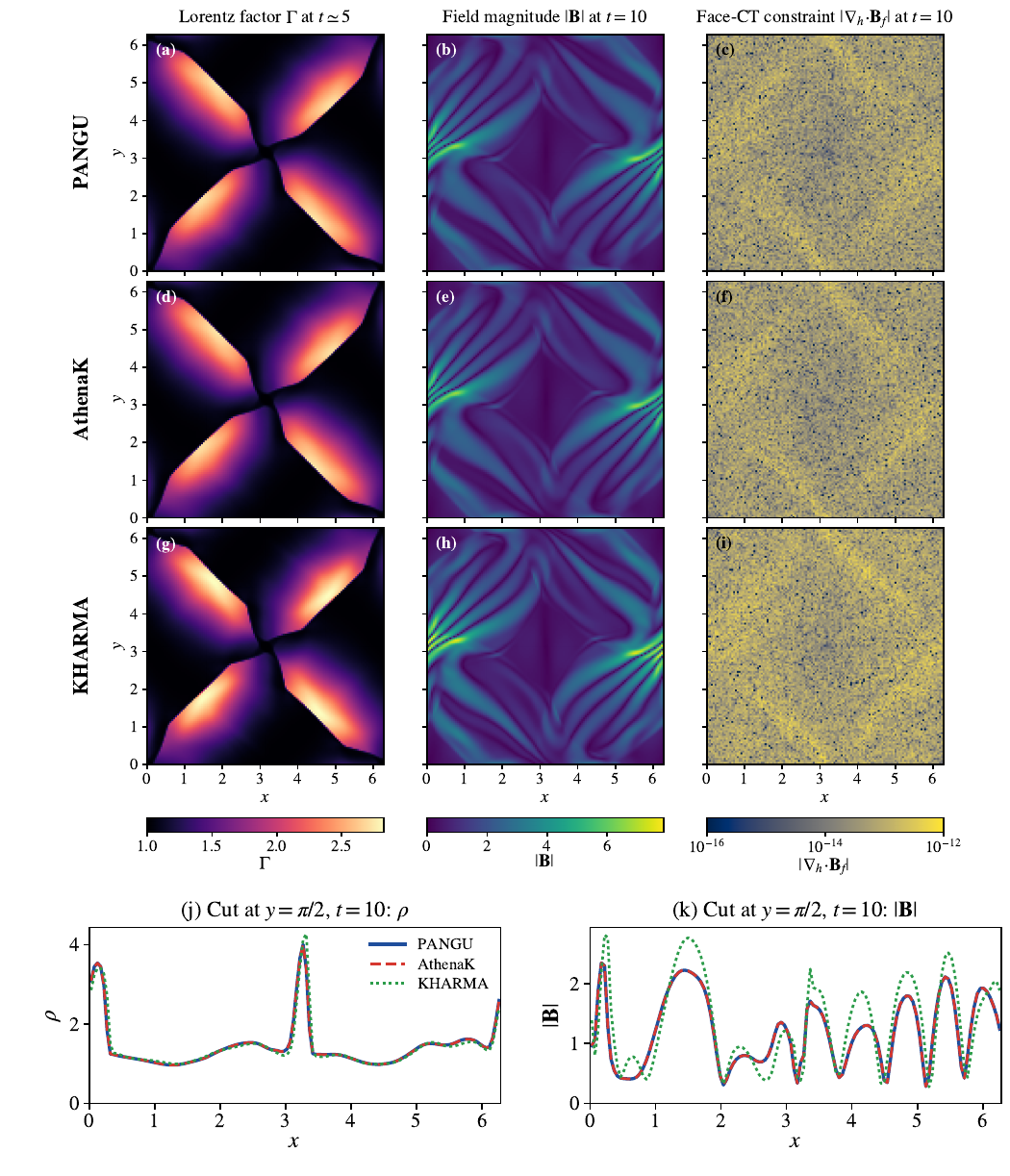}}
 \caption{Relativistic Orszag--Tang vortex on $128^2$ meshes.
     (a)--(i): rows show \pangu{}, \texttt{AthenaK}, and \texttt{KHARMA}; columns show the Lorentz factor at $t\simeq5$, the magnetic-field magnitude at $t=10$, and the unnormalized discrete face-centered-CT constraint at $t=10$.
   All panels in a column share one color scale.
    Exact zero divergence is displayed at the $10^{-16}$ lower color limit.
    (j), (k): horizontal cuts of $\rho$ and $|\bm B|$ at $y=\pi/2$ and $t=10$, linearly interpolated between the two adjacent cell rows, in the format of Figure~10 of \citet{Porth2017BHAC}.
    Across the three codes, the first two columns reproduce the same large-scale shock morphology, while the plotted discrete magnetic constraint remains at roundoff level.
    Positions and times are dimensionless.}
 \label{fig:ot}
\end{figure*}

\begin{figure*}[!tp]
  \centering
   \figframe{\includegraphics[width=0.98\textwidth]{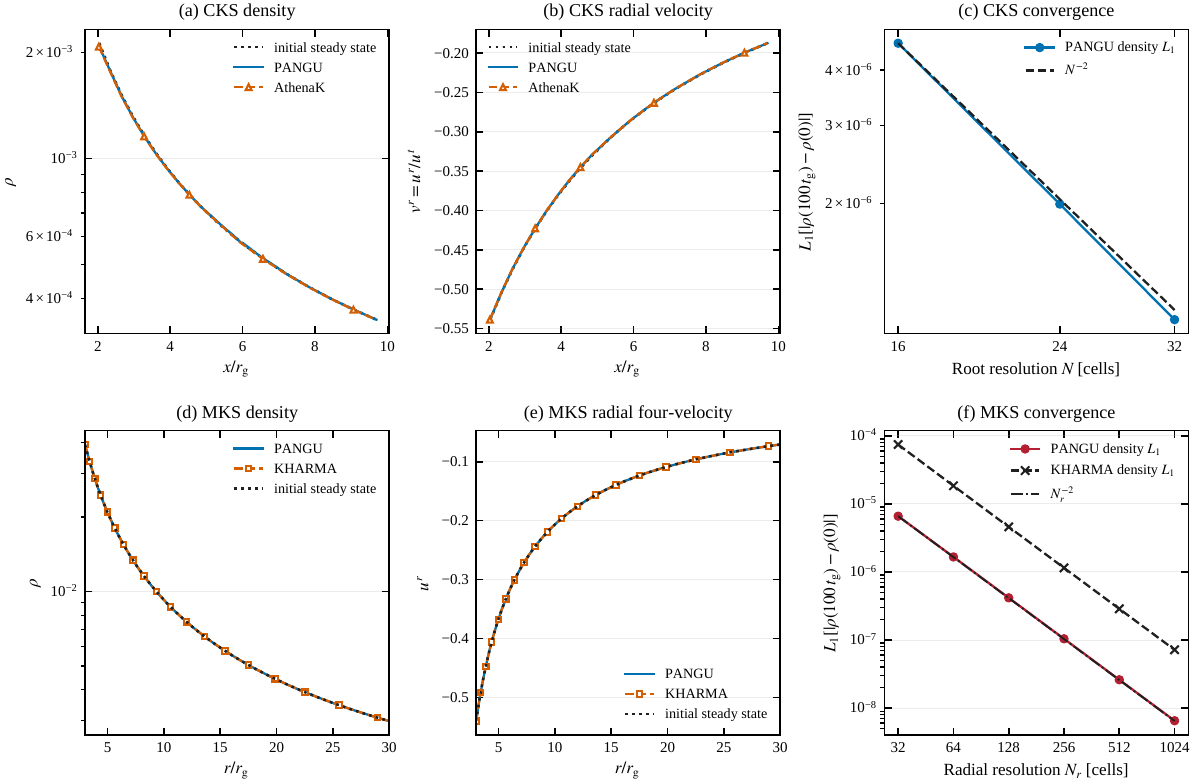}} \\[1.2ex]
   \figframe{\includegraphics[width=0.98\textwidth]{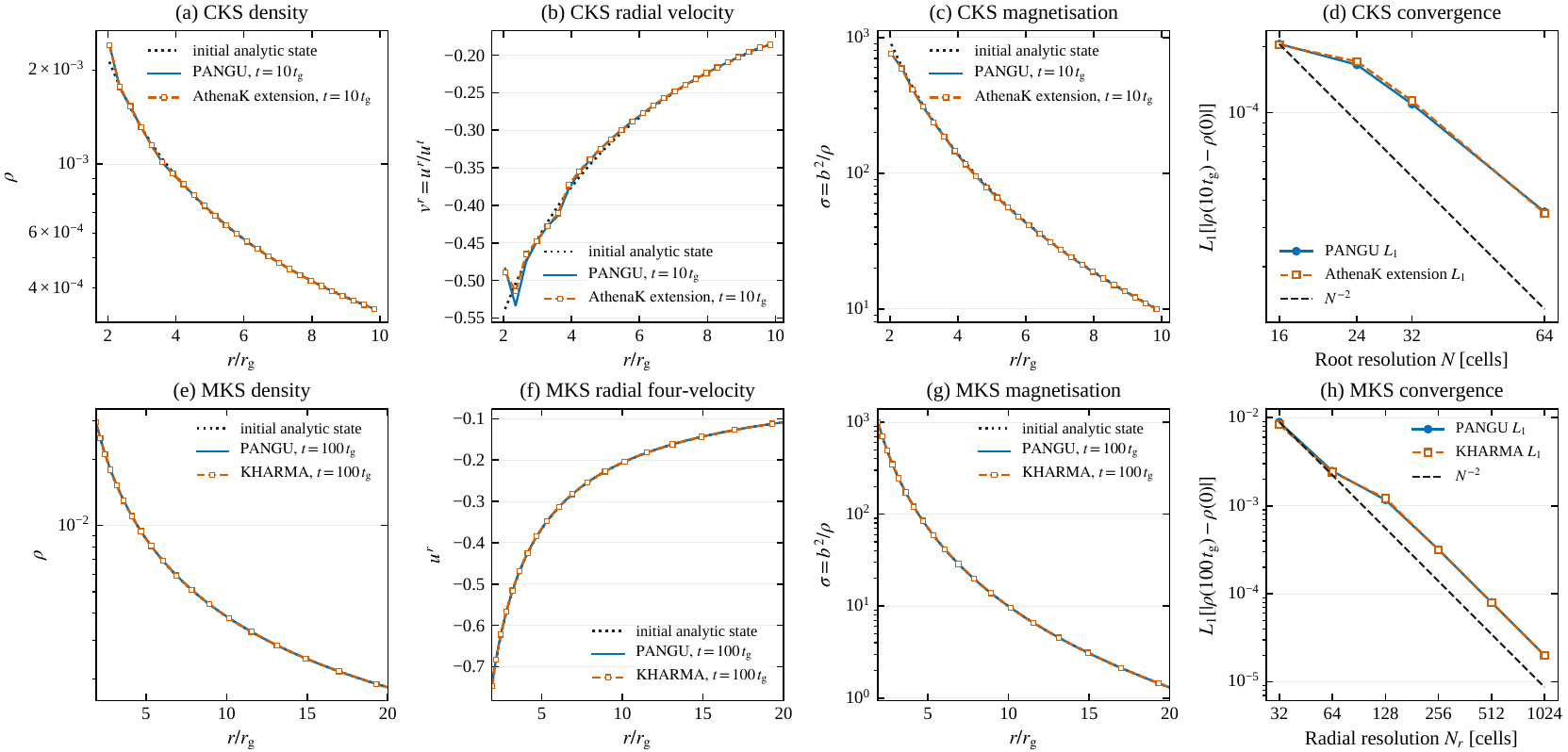}}
  \caption{Spherical-accretion verification.
   Upper: unmagnetized Bondi equilibrium in CKS/dynamic and MKS/static geometries, including native-grid convergence.
   Lower: the corresponding magnetized problem with initial 
   $\max\sigma=10^3$, including the magnetization profiles.
   CKS compares \pangu{} with \texttt{AthenaK}; MKS compares \pangu{} with \texttt{KHARMA}.
   Dotted curves are initial states, dashed lines are $N^{-2}$ references, and no native radial grid is interpolated.
   Together, the panels show stationary-flow preservation on the CKS/dynamic and MKS/static paths, while the magnetized case additionally stresses CT and primitive recovery.}
 \label{fig:bondi}\label{fig:magbondi}
\end{figure*}

The special-relativistic suite likewise separates smooth convergence from shock robustness and uses MUB1 as the matched SRMHD trajectory with its principal discretization settings stated below.
Five SRHD eigenfamilies are evolved for one crossing time at $N=32$--256.
The full-primitive $L_1$ orders are 1.59--1.86 between $N=32$ and 64, 1.96--2.02 between 64 and 128, and 2.02--2.10 between 128 and 256, where the five errors lie between $1.87\times10^{-10}$ and $2.42\times10^{-10}$.
Shock tubes and a two-dimensional relativistic blast separately test wave ordering, admissibility, and symmetry.
These tests distinguish smooth convergence from shock robustness rather than combining them into one fitted order \citep{MartiMuller2003,Font2008GRHydro, Pons1998GRRiemann,MignoneBodo2005HLLC,BeckwithStone2011RMHD}.

For SRMHD, the MUB1 problem \citep{Mignone2009MUB} uses 400 cells, PLM, HLLE, RK2, CFL 0.3, and six events to $t=0.4$   (514 cycles).
Over all events and primitive variables, the largest matched \pangu{}--\texttt{AthenaK} pointwise difference is $1.33\times10^{-14}$ (internal energy density, $\approx60\epsilon$), the largest $L_1$ difference is $1.11\times10^{-15}$, and $B_x$ is identical.
For comparison, \texttt{KHARMA} uses its native linear-MC reconstruction and flux-CT ordering, and differs from both codes by up to $0.104$ in $u_y$ (largest $L_1$ difference $2.53\times10^{-3}$).
This difference is localized at moving discontinuities, while the plateau states, wave ordering, and normal magnetic field agree, and both \pangu{} and \texttt{KHARMA} keep the stored divergence exactly zero.
Figure~\ref{fig:srmhd} therefore displays both types of evidence without claiming that unmatched shock widths should coincide.
An independent smooth SRMHD entropy wave at $N=32$--256 gives adjacent $L_1$ orders 1.86, 1.96, and 2.04 in \pangu{} and 1.66, 1.88, and 1.96 in \texttt{KHARMA}.

The relativistic Orszag--Tang problem further tests two-dimensional shocks and face-centered CT. \pangu{} and \texttt{AthenaK} recover the same Lorentz-factor and magnetic-field morphology, while \texttt{KHARMA} retains its native reconstruction and flux-CT path for an independent comparison of the shock pattern. The stored face-field constraint remains at roundoff level.
 At $t=10$ (619 \pangu{} cycles), the maximum unnormalized face-CT divergence is $4.70\times10^{-13}$ in \pangu{}, $5.35\times10^{-13}$ in \texttt{AthenaK}, and $7.70\times10^{-13}$ in \texttt{KHARMA}.
With $\Delta x=2\pi/128$ and $\max|\bm B|=6.72$, the \pangu{} value corresponds to $\max|\nabla\!\cdot\!\bm{\mathcal B}|\,\Delta x/\max|\bm B|=3.4\times10^{-15}$ ($\approx15\epsilon$).
 We do not replace this topological CT measure with a centered derivative of the averaged cell field, which can be large across shocks even when the discrete constraint is preserved.

Following \citet{Porth2017BHAC}, Fig.~\ref{fig:ot}(j,k) compares horizontal cuts of $\rho$ and $|\bm B|$ at $y=\pi/2$ and $t=10$.
Because $y=\pi/2$ lies on a cell face of the $128^2$ mesh, every code is sampled by linear interpolation between the two adjacent cell rows.
The \pangu{} and \texttt{AthenaK} cuts differ by at most $8.0\times10^{-8}$ in $\rho$ and $5.3\times10^{-8}$ in $|\bm B|$; this bound is set by the single-precision VTK output of \texttt{AthenaK}, not by the calculations.
\texttt{KHARMA} places the density peaks and field extrema at the same positions but reaches larger extrema, $\max\rho=4.27$ and $\max|\bm B|=2.81$ along the cut, compared with 4.01 and 2.35 in \pangu{}.
\citet{Porth2017BHAC} likewise find higher field maxima along this cut with their flux-interpolated CT than with GLM divergence cleaning.

\section{Fixed-spacetime GRMHD verification}

\begin{figure*}[!t]
  \centering
  \begin{minipage}[t]{0.495\textwidth}\centering
   {\large\textbf{A}}\\[0.35ex]\figframe{\includegraphics[width=0.94\linewidth]{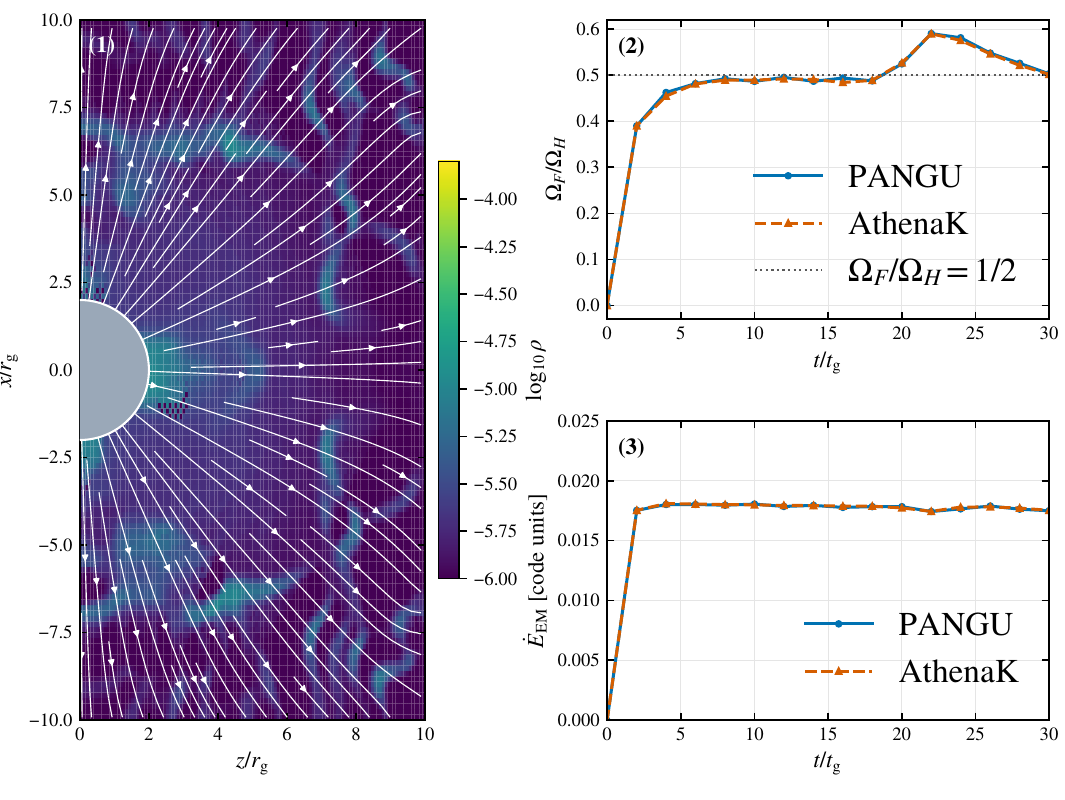}}
  \end{minipage}\hfill
  \begin{minipage}[t]{0.495\textwidth}\centering
   {\large\textbf{B}}\\[0.35ex]\figframe{\includegraphics[width=0.94\linewidth]{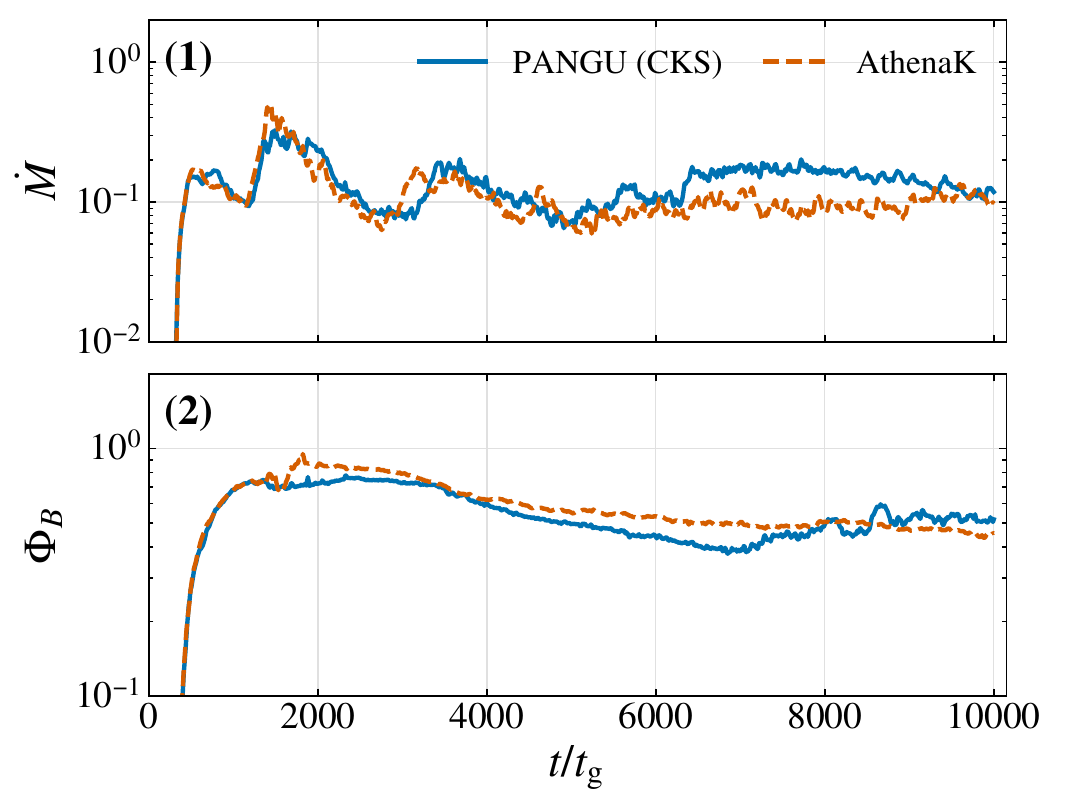}}
  \end{minipage}
  \begin{minipage}[t]{0.495\textwidth}\centering
   {\large\textbf{C}}\\[0.35ex]\figframe{\includegraphics[width=0.94\linewidth]{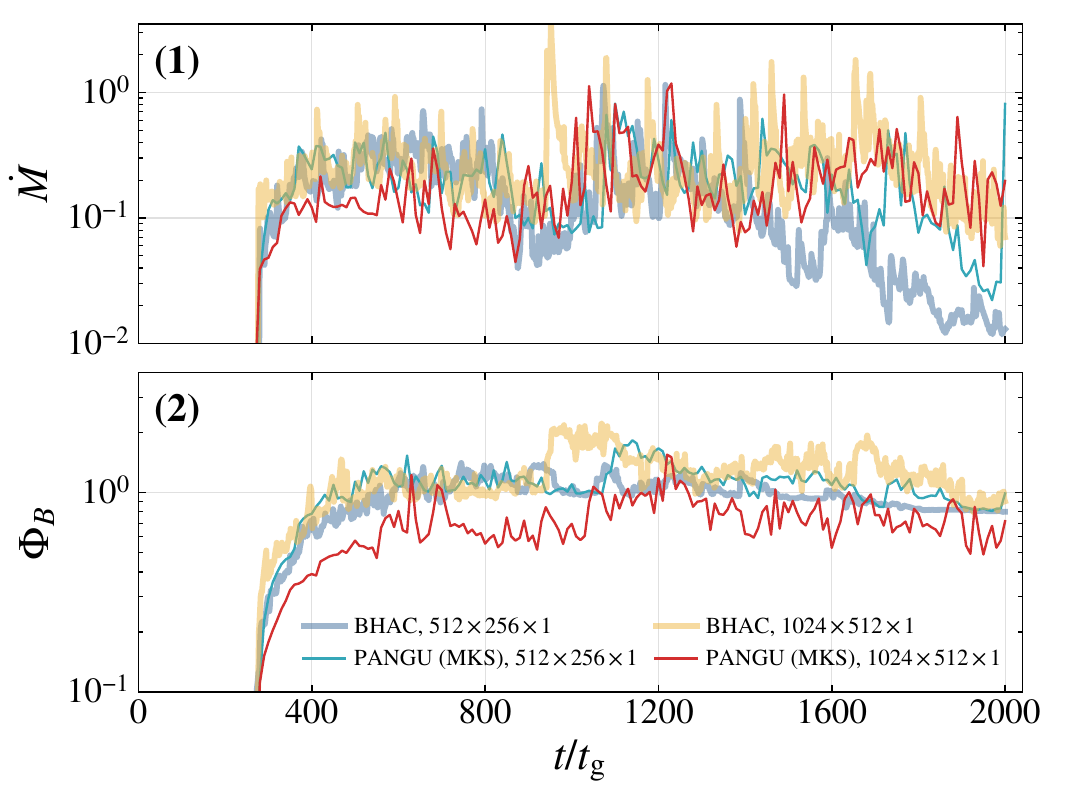}}
  \end{minipage}\hfill
  \begin{minipage}[t]{0.495\textwidth}\centering
   {\large\textbf{D}}\\[0.35ex]\figframe{\includegraphics[width=0.94\linewidth]{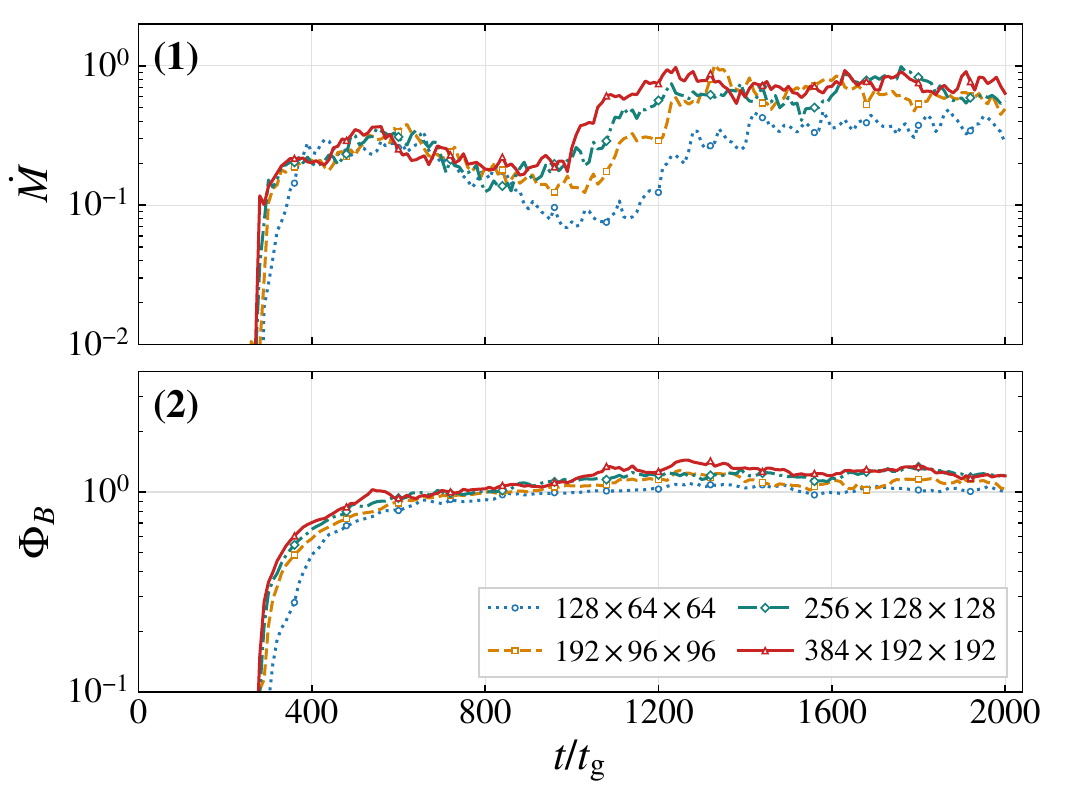}}
  \end{minipage}
  \caption{Fixed-spacetime black-hole diagnostics.
    \textbf{A}: CKS split monopole through $30\,\tg$ (panels 1--3: meridional state, field rotation, and outward electromagnetic energy flux).
    \textbf{B}: CKS/dynamic SANE horizon histories.
    \textbf{C}: axisymmetric MKS/static SANE comparison with \texttt{BHAC}.
    \textbf{D}: three-dimensional MKS/static resolution sequence.
    The SANE panels report $\dot M$ and unsigned horizon magnetic flux using Eq.~\eqref{eq:sane_diag}.
    Together, the panels provide horizon-scale split-monopole checks and show stable MRI-active accretion across both geometry paths; the SANE comparisons concern morphology and diagnostic scales rather than pointwise turbulent agreement.}
  \label{fig:monopole}\label{fig:cks_sane}\label{fig:mks_sane_2d}\label{fig:mks_sane_3d}
\end{figure*}

\begin{figure*}[!tp]
 \centering
 \figframe{\includegraphics[width=0.92\textwidth]{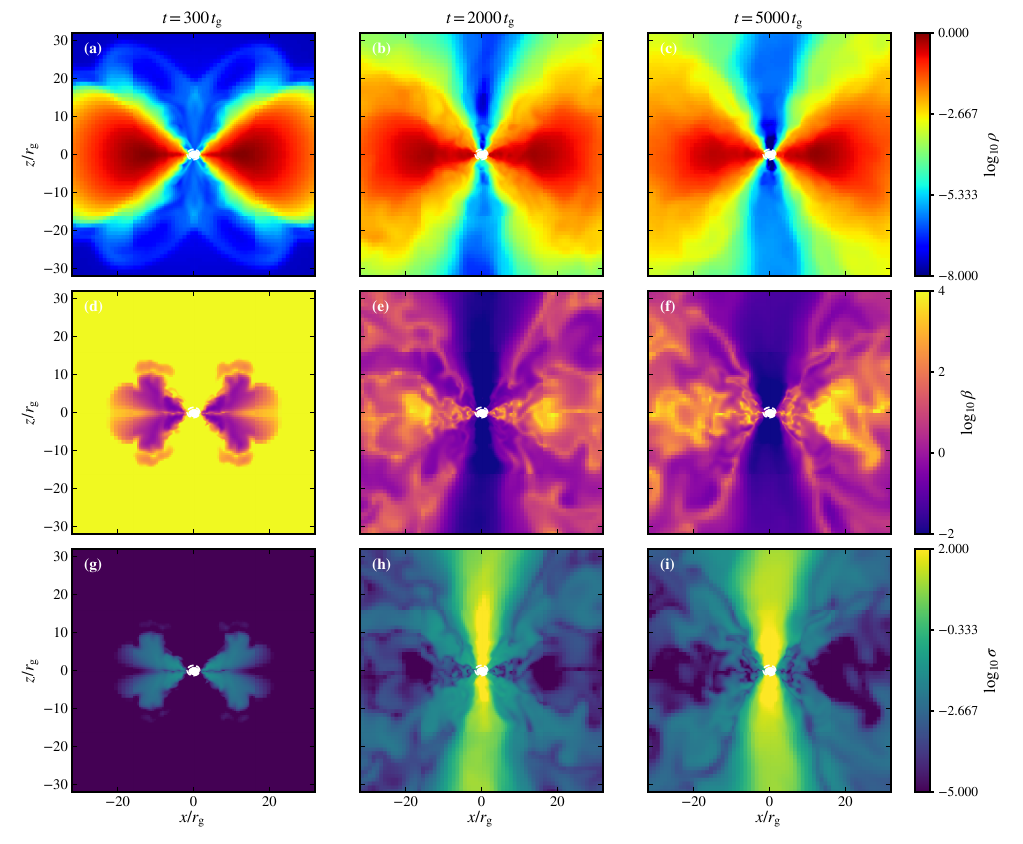}}
 \caption{CKS/dynamic SANE evolution on the native $y=0$ meridional SMR patches.
   The $64^3$ root grid uses $16^3$ MeshBlocks and four nested factor-two refinement levels, giving a finest cell spacing equivalent to a uniform $1024^3$ mesh.
   Columns show $t=300\,\tg$, $2000\,\tg$, and $5000\,\tg$; rows show $\log_{10}\rho$, $\log_{10}\beta$, and $\log_{10}\sigma$, respectively, each with one colorbar.
   The white dashed curve marks the event horizon.
   Display-only smoothing is applied independently within each native leaf patch.
   The sequence develops a dense disk, a magnetized corona, and a low-density polar funnel while retaining continuity across the displayed refinement levels.}
 \label{fig:cks_sane_states}
\end{figure*}

\begin{figure*}[!tp]
 \centering
   \figframe{\includegraphics[width=0.79\textwidth]{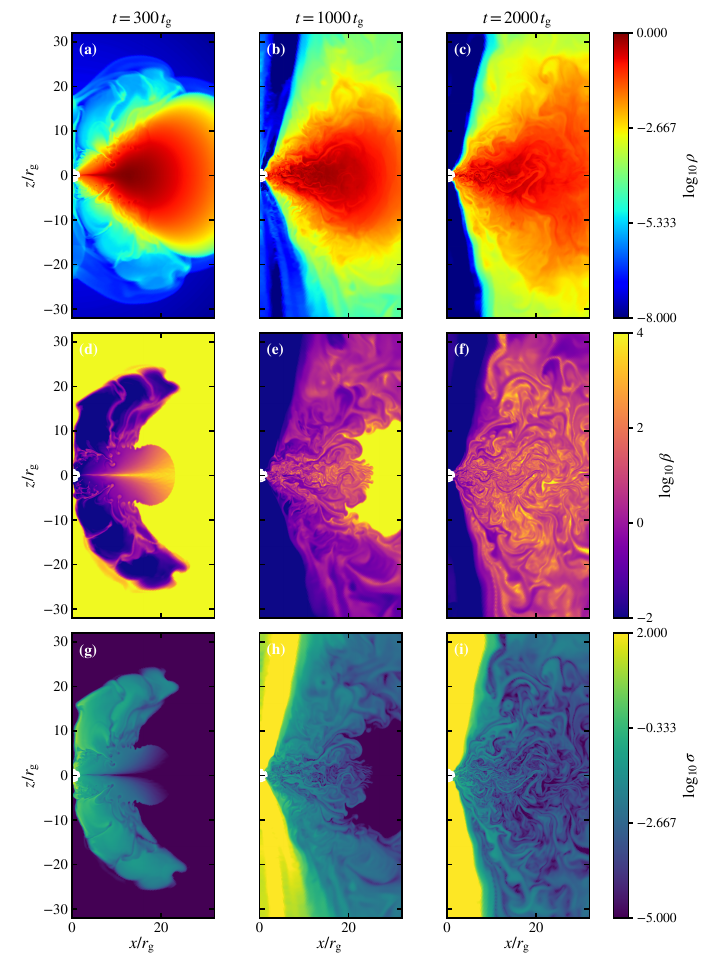}}
  \caption{MKS/static SANE evolution at $1024\times512\times1$ on the native axisymmetric mesh mapped to the positive-$x$ meridional half-plane.
   Columns show $t=300\,\tg$, $1000\,\tg$, and $2000\,\tg$; rows show $\log_{10}\rho$, $\log_{10}\beta$, and $\log_{10}\sigma$, respectively, each with one colorbar.
   Common limits are used in each row, and the white dashed curve marks the event horizon.
   The sequence shows the characteristic disk--corona--funnel structure and native-grid channel development during the early nonlinear axisymmetric evolution.}
\label{fig:mks_sane_states}
\end{figure*}
\label{sec:gr_results}

\subsection{Bondi accretion}

Relativistic spherical accretion supplies a stationary transonic solution \citep{Bondi1952,Michel1972}.
The boundary is geometry specific but not problem-specific within a black-hole topology: CKS applies analytic data on Cartesian physical faces, whereas MKS applies radial data on the two radial $X$ faces while polar parity and azimuthal periodicity are handled by the mesh.
Using a Cartesian six-face callback on MKS would therefore test the wrong boundary problem.

The CKS calculation uses a Schwarzschild background, a $32^3$ root mesh with one SMR region, and eleven states through $100\,\tg$.
With floors, Lorentz ceiling, CFL, and analytic sampling matched, the \pangu{}--\texttt{AthenaK} states are compared along the grid line $y=z=0.15625$ at all eleven states.
The largest primitive difference is $7.91\times10^{-16}$ (in $u^x$, $\approx3.6\epsilon$); density and pressure differ by at most $1.13\times10^{-17}$ and $3.14\times10^{-18}$, and the conserved variables by at most $1.91\times10^{-17}$.
A CKS resolution sequence with root meshes $N=16$, 24, and 32 reduces the density drift $L_1$ in the physical shell $3\,\rg\le r\le10\,\rg$ from $4.60\times10^{-6}$ to $1.09\times10^{-6}$, with orders 2.06 and 2.09.
The MKS sequence uses $N_r=32$--1024 and evolves to $100\,\tg$.
\pangu{}'s initial--final density $L_1$ drift decreases from $6.60\times10^{-6}$ to $6.50\times10^{-9}$, with adjacent orders 1.99--2.00.
At $N_r=1024$, the \pangu{}--\texttt{KHARMA} final $L_1$ differences are $6.53\times10^{-8}$ in density and $3.54\times10^{-9}$ in radial four-velocity.
Figure~\ref{fig:bondi} presents the profiles and convergence without interpolating native radial centers.
Together, the roundoff-level CKS trajectory agreement and the second-order CKS and MKS sequences show that both geometry paths preserve the stationary solution with their respective boundary implementations.

\subsection{Magnetized spherical accretion}

A radial magnetic field should not change the analytic Bondi fluid solution; it instead stresses CT, recovery, and the consistency of radial boundary data \citep{Porth2017BHAC}.
The initial field is scaled once so that the maximum $\sigma=b^2/\rho$ in the stated active shell is $10^3$.
This normalization is not an evolution-time ceiling.
The CKS $N=64$ final profiles have \pangu{}--\texttt{AthenaK} $L_1$ differences $2.42\times10^{-6}$ in density and $1.84\times10^{-3}$ in coordinate radial velocity; the relative $L_1$ difference in $\sigma$ is $2.13\times10^{-3}$.
The residual is concentrated near the Cartesian inner treatment rather than interpreted as magnetic modification of the analytic flow.

The MKS sequence uses $N_r=32$--1024, evolves to $100\,\tg$, and freezes the complete initialized GRMHD state in radial ghost zones.
At $N_r=1024$, the \pangu{}--\texttt{KHARMA} final differences are $4.61\times10^{-10}$ in density $L_1$, $1.54\times10^{-8}$ in radial-four-velocity $L_1$, and $3.64\times10^{-8}$ in relative magnetization $L_1$.
The density-drift $L_1$ decreases from $8.89\times10^{-3}$ to $2.00\times10^{-5}$ in \pangu{} and from $8.37\times10^{-3}$ to $2.00\times10^{-5}$ in \texttt{KHARMA}.
The adjacent orders are 1.84, 1.09, 1.87, 2.00, and 2.00 in \pangu{} and 1.80, 0.99, 1.94, 1.99, and 2.00 in \texttt{KHARMA}, so second order is reached for $N_r\ge256$.
The stored face-CT divergence is exactly zero at every history output and resolution in both codes,  and no recovery failure is hidden by a sigma ceiling.
The CKS and MKS results therefore test the coupled CT, primitive-recovery, and boundary implementations at high initial magnetization without invoking an evolution-time sigma ceiling.

\subsection{Split monopole}

The CKS split-monopole problem tests the horizon-regular, force-free limit and the Blandford--Znajek field rotation \citep{BlandfordZnajek1977, Komissarov2004Monopole}.
A $128^3$ calculation supplies 16 states at $2\,\tg$ cadence through $30\,\tg$.
Figure~\ref{fig:monopole}(A) locates the meridional state, field-rotation, and electromagnetic-energy-flux diagnostics for this test.
Outside the horizon and over all 16 states, the largest unnormalized \pangu{} face-CT divergence is $\max|\nabla\!\cdot\!\bm{\mathcal B}|=5.36\times10^{-14}$ on cells of width $\Delta x=20\,\rg/128$.
On the 8192 horizon samples at $t\simeq10\,\tg$, the four-velocity normalization error is $\max|u_\mu u^\mu+1|=1.22\times10^{-14}$, whose individual terms are of order $\gamma^2$ under the Lorentz ceiling $\gamma\le10$.
The field angular velocity is computed from the coordinate three-velocity obtained from Eq.~\eqref{eq:u4} and Appendix~\ref{app:primitive_velocity}, not by treating $\widetilde u^i$ as $u^i/u^0$.
\pangu{} and \texttt{AthenaK} approach $\Omega_F/\Omega_H\simeq1/2$ and develop the same outward electromagnetic energy flux.
Over the 16 stored events, the horizon median of $\Omega_F/\Omega_H$ differs between the codes by at most $3.89\times10^{-3}$ (mean $8.9\times10^{-4}$; final values 0.4697 and 0.4703), and the electromagnetic energy flux by at most $1.48\times10^{-4}$ (mean $5.9\times10^{-5}$), or 0.84\% of the final \pangu{} flux $1.75\times10^{-2}$.
Pointwise comparison inside the CKS shear layer is excluded.
These diagnostics therefore test the CKS horizon treatment, coordinate-velocity reconstruction, and electromagnetic-energy-flux calculation outside the excluded shear layer.

\subsection{SANE accretion}
\label{sec:sane}

\textbf{Setup and diagnostics.}
Fishbone--Moncrief tori supply the equilibrium state \citep{Fishbone1976Torus}.
We use a dimensionless Kerr spin $a=15/16$, $\Gamma=4/3$, an inner edge at $6\,\rg$, a pressure maximum at $12\,\rg$, and $\rho_{\max}=1$.
Poloidal loops are initialized from the unnormalized azimuthal vector potential
\begin{equation}
 A_\phi\propto\max\!\left(\frac{\rho}{\rho_{\max}}-0.2,\;0\right), \label{eq:sane_aphi}
\end{equation}
whose discrete curl supplies the face-centered field; the field is normalized so that $\max p/\max(b^2/2)=100$ over the torus and perturbed in pressure by 4\% to seed the MRI \citep{BalbusHawley1991MRI,HawleyGammieBalbus1995, DeVilliersHawley2003,McKinneyGammie2004,Narayan2012SANE}.

We monitor the horizon accretion rate and unsigned magnetic flux defined in Eq.~\eqref{eq:sane_diag}.
SANE evolution is an integrated stress test of metric-aware initialization, CT, primitive recovery, floors, boundaries, and restart semantics.
Because independently seeded MRI turbulence decorrelates individual events, we compare morphology and diagnostic scales rather than late-time pointwise values.

\textbf{CKS/dynamic path.}
The CKS run uses a $64^3$ root mesh, four nested SMR regions, PPM4, and FOFC.
Figure~\ref{fig:cks_sane}(B) follows \pangu{} and \texttt{AthenaK} through $10^4\,\tg$.
Both leave the weak-inflow phase, sustain positive accretion, and retain finite unsigned flux without runaway accumulation.
Individual peaks decorrelate, whereas their amplitudes and long-time variability remain in the same regime.

\textbf{MKS/static path.}
The axisymmetric MKS runs test logarithmic radial coordinates, polar boundaries, stored geometry, and wave-based time-step control.
In Fig.~\ref{fig:mks_sane_2d}(C), the $512\times256\times1$ and $1024\times512\times1$ \pangu{} runs reproduce the transition and diagnostic scales seen in independently reduced \texttt{BHAC} data.
Resolution changes the sharp channel episodes, as expected in nonlinear MRI evolution, but not the active accretion state or order-unity flux scale.

\textbf{Three-dimensional MKS sequence.}
The 3D campaign uses $128\times64\times64$, $192\times96\times96$, $256\times128\times128$, and $384\times192\times192$.
We restrict the comparison in Fig.~\ref{fig:mks_sane_3d}(D) to the common interval $0\leq t\leq2000\,\tg$ for all four resolutions.
Within this interval, the diagnostics show resolution-dependent onset and peak amplitudes, followed by MRI-active inflow and an order-unity horizon flux.
These early nonlinear trajectories characterize resolution sensitivity; the finite comparison window does not establish a statistically converged turbulent steady state.

The horizon diagnostics provide an integrated view of the three-dimensional sequence that is complementary to the meridional snapshots below.
The initial rise of $\dot M$ marks the establishment of inflow, while the subsequent variability reflects MRI-driven transport rather than a change in the metric or boundary prescription.
The magnetic-flux curves approach a common order-unity scale; their resolution-dependent fluctuations are therefore interpreted as trajectory sensitivity in a nonlinear flow.
We use these surface-integrated quantities to compare resolutions without treating decorrelated turbulent snapshots as pointwise matches.

\textbf{Morphology and interpretation.}
Figures~\ref{fig:cks_sane_states} and \ref{fig:mks_sane_states} show the corresponding density, $\beta=2p/b^2$, and $\sigma=b^2/\rho$ fields with fixed row-wise limits.
Both paths develop a dense disk, magnetized corona, and low-density polar funnel.
The CKS panels preserve native leaf patches and thus also expose inter-level continuity; the MKS panels expose channel structure on the native logarithmic mesh.
Axisymmetry limits the latter to early nonlinear MRI behavior rather than a sustained 3D dynamo \citep{HawleyGammieBalbus1995}.
Together, Figures~\ref{fig:cks_sane}--\ref{fig:mks_sane_states} show that the same GRMHD solver remains stable across the CKS/dynamic and MKS/static geometry paths.
They validate the integrated numerical chain, not a claim of pointwise turbulent agreement or a fully converged SANE steady state.

\section{Performance baseline}
\label{sec:performance-baseline}

\begin{figure}[!tb]
{\settablecaptype
 \caption{Median single-GPU baseline from seven independent processes.
  Throughput is in million active-cell updates s$^{-1}$; time is seconds per 100 cycles.}
\label{tab:performance}
\begin{ruledtabular}
 \setlength{\tabcolsep}{3pt}
 \begin{tabular}{llrrr}
Metric & Physics & \pangu{} & \texttt{AthenaK} & \texttt{KHARMA}\\
\hline
Throughput & HD & 19.936 & 23.602 & ---\\
Throughput & MHD & 11.752 & 11.643 & ---\\
Throughput & SRMHD & 8.533 & 6.652 & 3.511\\
Time/100 & HD & 1.315 & 1.111 & ---\\
Time/100 & MHD & 2.231 & 2.251 & ---\\
Time/100 & SRMHD & 3.072 & 3.941 & 7.466\\
\end{tabular}
\end{ruledtabular}
  }
\vspace{2ex}
  \centering
   \figframe{\includegraphics[width=0.95\columnwidth]{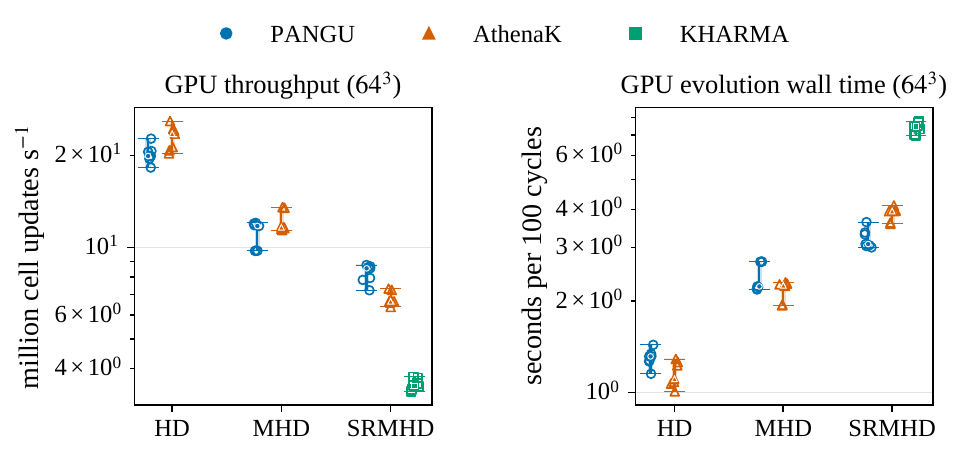}}
  \caption{Controlled single-GPU baseline from seven independent processes.
   Open symbols show individual measurements, filled symbols show medians, and bars span the observed range.
   On the tested RTX 3060 configuration, the medians show lower HD, comparable MHD, and higher SRMHD throughput for \pangu{} relative to the displayed comparison paths.}
\label{fig:performance}
  \end{figure}

Performance is evaluated separately from correctness.
The present baseline uses one NVIDIA GeForce RTX 3060 Laptop GPU, one $64^3$ block, FP64, RK2, and 100 measured cycles after ten warm-up cycles; initialization and output are excluded.
Seven fresh, interleaved processes are host-pinned with OpenMP off.
HD, MHD, and SRMHD compare \pangu{} with \texttt{AthenaK}, while \texttt{KHARMA} participates in SRMHD using its native reconstruction and CT path.
Figure~\ref{fig:performance} and Table~\ref{tab:performance} summarize the measurements.
A separate A100 MPI strong-scaling campaign evaluates magnetized Bondi accretion at fixed global size and is summarized from archived scheduler logs in Table~\ref{tab:multigpu}. Because this campaign uses different hardware, a different problem, and application-log timing, its throughput values are not directly comparable to the controlled RTX 3060 baseline.

  \begin{figure}[!tb]
 {\settablecaptype
 \caption{MPI strong scaling for the $3072\times1\times1$ magnetized Bondi problem through $1000\,\tg$.
  Every run reaches 2,289,417 cycles.
  Wall time $t_{\rm wall}$ and throughput (in $10^6$ cell-cycles s$^{-1}$) are the terminal \texttt{walltime used} and \texttt{zone-cycles/wallsecond} values reported by the application logs.
  Speedup $S$ is referenced to the one-GPU run on the same A100 memory class, and efficiency is $100S/N_{\rm GPU}$.}
\label{tab:multigpu}
\begin{ruledtabular}
  \footnotesize
\setlength{\tabcolsep}{3pt}
\begin{tabular}{lcrrrr}
 GPU type & $N_{\rm GPU}$ &   $t_{\rm wall}$  (s) & Throughput & $S$ &   Eff.  (\%)\\
\hline
 \multirow{4}{*}{A100 40 GB} & 1 &  5140 & 1.37 & 1.00 & 100.0\\
           & 2 &  3270 & 2.15 & 1.57 & 78.6\\
           & 3 &  2670 & 2.63 & 1.93 & 64.2\\
           & 4 &  2170 & 3.24 & 2.37 & 59.2\\
\hline
 \multirow{4}{*}{A100 80 GB} & 1 &  5190 & 1.35 & 1.00 & 100.0\\
           & 2 &  3290 & 2.14 & 1.58 & 78.9\\
           & 3 &  2670 & 2.64 & 1.94 & 64.8\\
           & 4 &  2180 & 3.23 & 2.38 & 59.5\\
  \end{tabular}
 \end{ruledtabular}
  }
\end{figure}

All seven repeats pass the output-checksum and timing-protocol checks.
Median throughputs are 19.936 and 23.602 million cell updates s$^{-1}$ for the \pangu{} and \texttt{AthenaK} HD paths, 11.752 and 11.643 for their MHD paths, and 8.533, 6.652, and 3.511 for \pangu{}, \texttt{AthenaK}, and \texttt{KHARMA} in SRMHD.
For these configurations, \pangu{} trails \texttt{AthenaK} in HD, gives comparable MHD throughput, and exceeds both comparison paths in SRMHD.
These measurements are a single-device baseline rather than a claim of architecture-independent ranking.

The two A100 memory classes give closely matched rates for each GPU count.
Across both allocations, the four-GPU configuration raises the terminal throughput from about $1.36\times10^6$ to $3.24\times10^6$ cell-cycles s$^{-1}$, corresponding to a strong-scaling speedup of $2.37$--$2.38$ and an efficiency of about 59\%.
These measurements establish an MPI execution baseline for this fixed global problem.
They do not extrapolate to a general hardware ranking or to larger per-GPU workloads, where communication and memory behavior must be measured separately.

\section{Conclusions and Discussion}
\label{sec:conclusion}\label{sec:discussion}

\pangu{} I provides unified Newtonian, SR, and fixed-spacetime GR HD/MHD capabilities through a \texttt{Parthenon} package architecture.
Within this architecture, the densitized GRMHD state, face-centered CT, scalar primitive recovery, local FOFC, and compile-time CKS/MKS geometry composition are defined and tested independently.
The evidence progresses from second-order convergence in smooth tests and FP64 roundoff-level matched trajectories, with maximum pointwise differences of $7.9\times10^{-16}$--$2.0\times10^{-14}$, to native cross-code convergence and integrated Bondi, monopole, and SANE accretion tests. Together, these results establish the fixed-spacetime foundation for the dynamical-spacetime formulation in Paper II.

The present conclusions have several defined limits.
Matched trajectories near FP64 roundoff jointly constrain initialization, state conversion, reconstruction, wave speeds, Runge--Kutta stages, CT orientation, boundaries, and output semantics only for deliberately matched discretizations.
Native \texttt{KHARMA} shock and MKS comparisons retain different operators and therefore establish a common convergent solution class rather than identical cells at discontinuities.
The two-dimensional MKS SANE runs probe early MRI development, and the three-dimensional resolution sequence is restricted to $0\leq t\leq2000\,\tg$; neither supports long-time turbulent statistics.
Optional electron thermodynamics is available; implementation and regression details are documented in the \pangu{} Wiki.
The RTX 3060 timing baseline excludes communication and I/O, whereas the A100 measurements establish MPI strong scaling only for the fixed global magnetized Bondi problem in Table~\ref{tab:multigpu}.
Radiation transport and dynamical-spacetime evolution remain outside the scope of Paper I.

These limits clarify how the validated geometry paths should be used.
MKS/static is advantageous for two-dimensional black-hole calculations because its axisymmetric metric fields are stored once and reused.
In three dimensions, its wave-based estimator and polar coordinate angular speed can drive the time step toward a small, flow-dependent plateau.
CKS/dynamic avoids stored geometry fields and admits Cartesian excision, SMR, and light-based time-step estimation; it is therefore the preferred refined three-dimensional path in the present implementation, at the cost of recomputing metric quantities at their points of use.
The validated CKS production recipe combines SMR, PPM4, and FOFC.

The absence of refinement in static MKS is a current compute--memory design choice, not a mathematical restriction on stationary metrics.
A fully covered three-dimensional hierarchy with $L$ factor-of-two refinement levels could increase stored geometry data by as much as $2^{3L}$, whereas dynamic geometry exchanges this storage cost for arithmetic.
Because geometry selection remains orthogonal to the finite-volume solver and optional packages remain orthogonal to geometry, future work can address this tradeoff without restructuring the matter update.
The present evidence boundaries motivate distinct extensions through longer turbulent ensembles, communication-inclusive scaling, active electron thermodynamics, radiation transport, and dynamical-spacetime coupling.
One planned extension is to couple \pangu{} to the covariant polarized radiative transfer code \texttt{Coport} \citep{Huang2024Coport,Zhou2026FluxEruptions} to generate synthetic polarized images directly from \pangu{} GRMHD outputs.
Tilted-disk and extended electron examples are documented in the \pangu{} Wiki, \url{https://github.com/adamdarx/PANGU/wiki}, following the relevant equilibrium and alignment literature \citep{Chakrabarti1985Torus,Liska2019BPAlignment,Fragile2024TiltedReview}.

\begin{acknowledgments}
The development and validation of \pangu{} benefited from the published methods and openly available implementations of \texttt{Parthenon}, \texttt{Kokkos}, \texttt{AthenaK}, \texttt{KHARMA}, \texttt{HARMPI}, and \texttt{BHAC}.
We thank Yosuke Mizuno, Ye Shen, Indu Kalpa, Hongxuan Jiang, and Akhil Uniyal for useful discussions.
During our use of the \texttt{Parthenon} framework, Y.L. gratefully thanks Jonah Miller at LANL and Philipp Grete at the University of Hamburg for helpful suggestions and debugging strategies.
This work was partly supported by NSFC Grant Nos. 12575048, 12275004, and 12588101.
M.G. also acknowledges support from the BNU Tang Scholar Program.
This work is also supported by the high-performance computing platform of Peking University.
\end{acknowledgments}

\software{
  \pangu{} (DOI: \href{https://doi.org/10.5281/zenodo.22857644}{\texttt{10.5281/zenodo.22857644}}),
  \texttt{Parthenon} \citep{Grete2023Parthenon},
  \texttt{Kokkos} \citep{Edwards2014Kokkos,Trott2022Kokkos},
  \texttt{AthenaK} \citep{Stone2024AthenaK},
  \texttt{KHARMA} \citep{Prather2024KHARMA},
  \texttt{BHAC} \citep{Porth2017BHAC}
}

\par\vspace{1.2\baselineskip}
\section*{Code and data availability}

The version of \pangu{} used in this work is available at \url{https://github.com/adamdarx/PANGU} and archived on Zenodo at \href{https://doi.org/10.5281/zenodo.22857644}{DOI: 10.5281/zenodo.22857644}.
The relevant data are available from the authors upon reasonable request.

 \appendix
\renewcommand{\theHequation}{\thesection.\arabic{equation}}
\section{Primitive-velocity parametrization}
\label{app:primitive_velocity}

A direct primitive-velocity choice would use the spatial coordinate components $u^i$.
For prescribed $u^i$, however, four-velocity normalization gives
\begin{equation}
 g_{00}(u^0)^2+2g_{0i}u^iu^0+g_{ij}u^iu^j+1=0,                 \label{eq:u0_quadratic_appendix}
\end{equation}
which is a quadratic equation for $u^0$.
Requiring a real future-directed root restricts the admissible domain of $u^i$ in a metric-dependent manner, which is inconvenient to preserve during numerical reconstruction.

Following \citet{McKinneyGammie2004}, we instead use the Eulerian-frame spatial four-velocity $\widetilde u^i$.
For a foliation by $t={\rm const.}$ hypersurfaces, the metric takes the 3+1 form \citep{Arnowitt2008,Porth2017BHAC}
\begin{equation}
 \dd s^2=-\alpha^2\dd t^2
 +\gamma_{ij}(\dd x^i+\beta^i\dd t)(\dd x^j+\beta^j\dd t),   \label{eq:31_appendix}
\end{equation}
where $\alpha=(-g^{00})^{-1/2}$, $\beta^i=\alpha^2g^{0i}$, $\gamma_{ij}=g_{ij}$, and $\sqrt{-g}=\alpha\sqrt{\gamma}$.
The Eulerian observer normal to these hypersurfaces has four-velocity $n^\mu$, and the fluid four-velocity is decomposed as
\begin{equation}
 \begin{aligned}
 n^\mu&=\frac{1}{\alpha}(1,-\beta^i),
 &u^\mu&=W(n^\mu+v^\mu),\\
 \widetilde u^i&\equiv Wv^i,
 &W&=\sqrt{1+\gamma_{ij}\widetilde u^i\widetilde u^j},
 \end{aligned}                                                   \label{eq:utilde_appendix}
\end{equation}
where $v^\mu=(0,v^i)$ is the fluid three-velocity measured by the Eulerian observer and $W$ is its Lorentz factor.
Because $\gamma_{ij}$ is positive definite, $1+\gamma_{ij}\widetilde u^i\widetilde u^j\geq1$ for every $\widetilde{\bm u}\in\mathbb R^3$.
Thus $W$ is real throughout the unconstrained primitive-velocity domain, and the complete future-directed four-velocity follows from
\begin{equation}
 u^0=\frac{W}{\alpha},\qquad
 u^i=\widetilde u^i-\frac{W}{\alpha}\beta^i.                  \label{eq:u4_appendix}
\end{equation}

\section{Supported stationary metrics}
\label{sec:metrics}
The geometry interface supplies the metric and its derived 3+1 quantities at cell centers and at the face or edge locations required by the discretization.
This work exercises three stationary representations.

Special-relativistic calculations use Cartesian Minkowski coordinates,
\begin{equation}
 \dd s^2=-\dd t^2+\dd X^2+\dd Y^2+\dd Z^2,
 \label{eq:minkowski_metric_appendix}
\end{equation}
for which $g=-1$, $\alpha=1$, and $\beta^i=0$.

Cartesian Kerr--Schild (CKS) coordinates write the Kerr metric as the rank-one perturbation
\begin{align}
 \dd s^2={}&\eta_{\mu\nu}\dd X^\mu\dd X^\nu
 +2H(\ell_\mu\dd X^\mu)^2,\qquad
 H=\frac{r^3}{r^4+a^2Z^2},                                    \label{eq:cks_metric_appendix}\\
 \ell_\mu\dd X^\mu={}&\dd t+\frac{rX+aY}{r^2+a^2}\dd X
 +\frac{rY-aX}{r^2+a^2}\dd Y+\frac{Z}{r}\dd Z.\nonumber
\end{align}
Here $a$ is the dimensionless Kerr spin parameter, and
\begin{equation}
 r^2=\frac{1}{2}\left[R^2-a^2+\sqrt{(R^2-a^2)^2+4a^2Z^2}\right],
 \qquad R^2=X^2+Y^2+Z^2 .
\end{equation}
Because $\ell_\mu$ is null with respect to $\eta_{\mu\nu}$, the determinant is constant, $g=-1$.
The horizon-penetrating Cartesian chart has no polar axis and bounds each coordinate component of a null ray by unity.

These properties make on-demand CKS geometry effective for three-dimensional SMR calculations; the current CKS path is three-dimensional only \citep{Kerr1963,Gammie2003HARM,Stone2024AthenaK}.

Modified Kerr--Schild (MKS) coordinates begin with the spherical ingoing Kerr--Schild line element
\begin{align}
 \dd s^2={}&-\left(1-\frac{2r}{\Sigma}\right)\dd t^2
 +A\dd r^2+\Sigma\dd\theta^2
 +\sin^2\theta\left(\Sigma+a^2A\sin^2\theta\right)\dd\phi^2\nonumber\\
 &+\frac{4r}{\Sigma}\dd t\dd r
 -\frac{4ar\sin^2\theta}{\Sigma}\dd t\dd\phi
 -2aA\sin^2\theta\dd r\dd\phi,                               \label{eq:mks_metric_appendix}\\
 \Sigma={}&r^2+a^2\cos^2\theta,\qquad A=1+\frac{2r}{\Sigma}.\nonumber
\end{align}
The modified coordinates apply the map
\begin{equation}
 r=e^X,\qquad
 \theta=\pi Y+\frac{1-h_{\rm MKS}}{2}\sin(2\pi Y),\qquad \phi=Z .
 \label{eq:mks_map_appendix}
\end{equation}
where $h_{\rm MKS}$ controls the concentration of grid cells toward the equatorial plane.
The determinant varies spatially, $\sqrt{-g}=|\Sigma\sin\theta\,r\,\dd\theta/\dd Y|$, while axisymmetry removes the $Z$ extent from stored geometry.
Logarithmic radial spacing, equatorial concentration, and symmetry-reduced storage make MKS particularly efficient in two dimensions.

\subsection{Coordinate dependence of CFL estimates}
\label{sec:cfl_coordinates}

Equation~\eqref{eq:cflwave} has the same structure in all relativistic configurations. Its coordinate signal speeds depend on the metric and coordinate chart, while its coordinate cell widths depend on the grid mapping.

\textit{Cartesian Minkowski coordinates.} No metric evaluation is required. Both estimators use relativistic acoustic or fast-magnetosonic characteristic speeds. The \texttt{light} path takes their directional maximum, whereas \texttt{wave} uses the multidimensional reciprocal sum in Eq.~\eqref{eq:cflwave}.

\textit{CKS coordinates.} Because every coordinate component of a null ray is bounded by unity, \texttt{light} may use $\Delta t=C_{\rm CFL}\min_d\Delta x_d$ \citep{Stone2024AthenaK}. This state-independent estimate avoids metric and characteristic evaluations and is useful for long three-dimensional Cartesian SMR evolutions, at the cost of ignoring locally slower fluid waves. The alternative \texttt{wave} path evaluates metric-dependent characteristics, bounds them by the coordinate light cone, and applies Eq.~\eqref{eq:cflwave}.

\textit{MKS coordinates.} The nonuniform map and polar geometry allow coordinate signal speeds greater than unity, so MKS requires \texttt{wave} \citep{Gammie2003HARM,Prather2024KHARMA}. \pangu{} applies Eq.~\eqref{eq:cflwave} to metric-dependent fast-wave speeds and reuses the one-sided face speeds from the Riemann solver. In two dimensions, axisymmetry removes the $\phi$ term. In three dimensions, small proper azimuthal cells and increasing angular speed near the hole can make the azimuthal term dominate the CFL sum, producing a small, fluctuating global time step.

  \bibliographystyle{aasjournalv7}
 \bibliography{references}

\end{document}